\documentclass[a4paper,12pt]{article}
\usepackage{amsmath,amssymb,amsthm,bm,empheq,extarrows}
\usepackage[paper=letterpaper,margin=1.0in]{geometry}
\usepackage{cite}
\usepackage{hyperref}
\usepackage{graphicx}
\usepackage{subfig}

\graphicspath{{./fig/}}
\newcommand{\be}{\begin{equation}}
\newcommand{\ee}{\end{equation}}
\newcommand{\bea}{\begin{eqnarray}\displaystyle}
\newcommand{\eea}{\end{eqnarray}}
\newcommand{\bi}{\begin{itemize}}
\newcommand{\ei}{\end{itemize}}
\newcommand{\bfig}{\begin{figure}}
\newcommand{\efig}{\end{figure}}
\newcommand{\ret}{\nonumber \\}

\newcommand{\eref}[1]{(\ref{#1})}

 \def\cB{{\cal B}} 
  
\def\cG{{\cal G}}  
  
\def\cM{{\cal M}}  
  
 \def\cT{{\cal T}}

\newcommand{\cor}[1]{\langle{#1}\rangle}
\def\AG{\mathrm{Aut}(\cG)}
\def\Afix{\mathrm{Aut}_{\mathrm{fix}}(\cG)}
\def\afix{a_{\mathrm{fix}}}

\begin{document}

\rightline{QMUL-PH-15-14}

\vspace*{2cm} 

{\LARGE{  \bf 
\centerline{The  geometry of the light-cone   } } 
{ \LARGE \bf  \centerline  {cell decomposition of  moduli space } } }

\vskip.5cm

\thispagestyle{empty} \centerline{
    {\large \bf David Garner 
\footnote{ {\tt d.p.r.garner@qmul.ac.uk}}}
   {\large \bf and Sanjaye Ramgoolam
               \footnote{ {\tt s.ramgoolam@qmul.ac.uk}}   }
}   
               
\vspace{.4cm}

\vspace{.4cm}

\centerline{{\it Centre for Research in String Theory,}}
\centerline{ {\it School of Physics and Astronomy},}
\centerline{{ \it Queen Mary University of London},} 
\centerline{{\it    Mile End Road, London E1 4NS, UK}}

\vspace{1.4truecm}

%%%%%%%%%%%%%%%%%
\thispagestyle{empty}

\centerline{\bf ABSTRACT}
\vspace*{1cm}

The moduli space of Riemann surfaces with at least two punctures can be decomposed into a cell complex
by using a particular family of ribbon graphs called Nakamura graphs. 
We distinguish the moduli space with all  punctures labelled from that with a single labelled  puncture. 
In both cases, we describe a cell decomposition where the cells are parametrised by graphs or  
equivalence classes of finite sequences (tuples) of permutations. 
Each cell  is a convex polytope   defined by a system of 
linear equations and inequalities relating light-cone string parameters,  quotiented by the automorphism group of the graph. 
We give explicit examples of the cell decomposition at low genus with few punctures.  
% , and prove that the automorphism group of a cell
% coincides with the automorphism group of the graph.

\vskip.4cm

\setcounter{page}{0}
\setcounter{tocdepth}{2}

\newpage

\tableofcontents

\section{Introduction}

The moduli space of inequivalent Riemann surfaces is of fundamental interest to mathematicians and theoretical physicists.
The study of Riemann surfaces, mapping class groups, and Teichm\"uller theory lies at the intersection of group theory, algebraic geometry, and the theory of complex manifolds. It is also an essential ingredient in the most basic formulations of string theory, 
arising in the integration measure of bosonic string amplitudes.
However, all but the simplest moduli spaces are topologically very complicated, and difficult to describe explicitly.

One approach to understanding the topology of moduli space is to use the {\it light-cone cell decomposition} of moduli space. In this decomposition, each point in moduli space has an associated {\it light-cone diagram} consisting of several glued parallel cylinders, and in which the continuous parameters of a Riemann surface are encoded into the widths and lengths of the cylinders, and in the `twist' parameters of their gluings.
Giddings and Wolpert showed in \cite{gw} that each closed string light-cone diagram corresponds to a punctured Riemann surface with a uniquely-determined meromorphic one-form, later called the {\it Giddings-Wolpert differential}.
This correspondence was used to argue that the light-cone string diagrams lead to a single cover of moduli space, which is necessary for the light-cone and covariant formulations of string theory to be equivalent. However, this approach had some issues involving the overcounting of discrete factors, analogous to the symmetry factors appearing in Feynman diagram expansions. In addition, the higher order string interactions are tricky to describe in the light-cone picture.

These issues were addressed by Nakamura in \cite{nakamura} with the introduction of a particular type of ribbon graph on each punctured Riemann surface determined by the Giddings-Wolpert differential.
Every Riemann surface has a uniquely-determined embedded graph of this type called the {\it Nakamura graph} of the surface.
The vertices of this graph correspond to the punctures of the Riemann surface and the interaction points of a light-cone diagram. The faces of this graph are holomorphic to strips, with the points at infinity corresponding to the punctures of the Riemann surface, and the interaction points living on the finite boundaries of the strips.
The set of distinct Nakamura graphs partitions moduli space into disjoint convex polytopes, with each polyhedron parametrised by the relative positions of the interaction points and the widths of the strips of each surface.
These polytopes give a cell decomposition of moduli space.
The validity of this cell decomposition was shown computationally in \cite{nakamura} by counting the distinct graphs for low genus and few punctures, evaluating the orbifold Euler characteristic of the decomposition, and comparing this to known exact formulae from \cite{hz}.
This is highly non-trivial evidence for the consistency of the {light-cone cell decomposition}, as a very large number of graphs were counted to confirm the orbifold Euler characteristic.

Recently, this graphical approach to cataloguing the cells of moduli space was reconsidered in \cite{fgr}.
It was shown that Nakamura graphs have several descriptions in terms of equivalence classes of permutation tuples,
which arise from links between Nakamura graphs, Grothendieck's dessins d'enfants \cite{grothendieck}, and branched coverings of the sphere. This approach leads to links between the counting of cells of moduli space and Gaussian matrix models, and also leads to powerful new methods of calculating topological invariants of moduli space computationally. 
The orbifold Euler characteristics of moduli spaces $\cM_{g,n}$ with $2g+n=9$ were calculated by counting Nakamura graphs via permutation tuples, providing even stronger evidence for the validity of the cell decomposition of moduli space via Nakamura graphs.
This calculation involved the cell decomposition of a slightly modified version of moduli space, in which $(n-1)$ of the $n$ labelled points are treated as indistinguishable.

In this paper, we develop this description of cells of moduli space in terms of Nakamura graphs and Hurwitz equivalence classes. We clarify the relations between the modified moduli space, which we denote $\cM_{g,1[n-1]}$, and the conventional moduli space $\cM_{g,n}$.
The recent paper \cite{fgr} considered several ways to describe Nakamura graphs in terms of permutation tuples, and the relations between them; in this paper, we focus on the description arising from branched coverings of the sphere, called the {\it $S_d$ description}. We extend this tuples description of cells in the modified moduli space $\cM_{g,1[n-1]}$ to a tuples description of cells in the conventional moduli space $\cM_{g,n}$ by introducing a type of permutation tuple called a {\it split tuple}.
From this new type of tuple, we can show that the automorphism group of a Nakamura graph coincides with the orbifold group of its associated cell in moduli space, as was implicitly assumed in the graph-counting calculations of \cite{nakamura, fgr}.

The split permutation tuple description of Nakamura graphs leads to an algorithmic way of constructing a cell and finding its boundaries and orbifold quotienting group.
From the structure of each permutation tuple, we can find an associated system of linear equations specifying a real convex polyhedron. 
The parameters specifying this polyhedron correspond to the widths of the strips and the relative positions of the interaction times of a Riemann surface with an embedded Nakamura graph.
The orbifold quotienting group acts on this polyhedron by interchanging these parameters, and the quotient space is in one-to-one correspondence with a cell in moduli space.
We demonstrate this procedure by providing some simple explicit examples of moduli spaces and their Nakamura graph cell decompositions, and showing that they coincide.
Finally, as a way of showing explicitly the invariance of the Nakamura graphs cell decomposition under the mapping class group, we discuss an extension of the Nakamura graphs description of moduli space to a description of Teichm\"uller space.

\section{Review}
We give a quick review of the Giddings-Wolpert differentials, their link to Nakamura graphs and strip decompositions of 
Riemann surfaces, as well as the permutation-tuple description of the graphs \cite{gw, nakamura, fgr}.

\subsection{Giddings, Wolpert, and Nakamura}

Consider a connected Riemann surface $X$ with $n$ distinguished labelled points $P_1, P_2, \ldots, P_n$ and genus $g$, where $n\geq 2$.
Associate a set of real numbers $r_1, r_2, \ldots r_n$ respectively to the $n$ labelled points, which satisfy $\sum_i r_i=0$.
Giddings and Wolpert proved in \cite{gw} that there exists a unique abelian differential $\omega$
on the Riemann surface $X$ such that $\omega$ has $n$ simple poles at the points $P_i$ with respective residues $r_i$ and pure imaginary periods on any closed integral on the surface.
This is the {\bf Giddings-Wolpert} differential of the surface.
This differential restricts to a holomorphic differential on $\hat{X}$, the punctured Riemann surface with the $n$ labelled points removed.

The Giddings-Wolpert differential yields a global time coordinate on the Riemann surface $X$, up to an overall constant representing the time translation symmetry. 
If we fix a point $z_0$ on the surface which is not a pole of $\omega$, 
then we can define the global time coordinate of a generic point $z$ on the surface to be $T := \mathrm{Re}(\int_{z_0}^z \omega)$.
This expression does not depend on the choice of integration contour from $z_0$ to $z$, since any two paths from 
$z_0$ to $z$ differ only by a closed contour, and the integral of the differential along any closed contour is imaginary.
The global time coordinate tends to negative infinity as we approach the poles with positive residue, and to positive infinity as we approach the poles with negative residues.
We call the poles with positive residue the {\bf incoming poles}, and the poles
with negative residue the {\bf outgoing poles}.
Examples of explicit constructions of Giddings-Wolpert differentials on surfaces are given in \cite{gw, fgr}.

At any point which is not a pole of the Giddings-Wolpert differential, it is possible to choose local complex coordinates about the point such that $\omega = d(z^{p+1})$ for some non-negative integer $p$. The set of zeroes of the Giddings-Wolpert differential is the set of points 
$Q_1, \ldots, Q_l$ on the surface at which $p$ is positive. The integer $p$ is the {\bf order} of the zero: if $p=1$, the zero is {\bf simple}.
For each point on the surface which is not a pole of $\omega$, there exists a finite set of directions in which $z^{p+1}$ is real. These are the {\bf real trajectories} that extend out from the point.
There are two real trajectories extending out from a generic point on the surface,
and $2(p+1)$ real trajectories extending out from a zero of order $p$.
The real trajectories that extend out from the zeroes of the differential will only meet at the poles and zeroes of the differential.

The set of real trajectories extending out from all the zeroes of the differential defines a  ribbon graph embedded onto the surface, 
with the vertices of the graph corresponding to the poles and zeroes of $\omega$, and the edges of the graph corresponding to the real trajectories extending out from zeroes.
The edges also inherit an orientation from the Giddings-Wolpert differential: each edge is oriented in the direction along which the global time coordinate increases.
It was shown in \cite{nakamura} that this graph has the following properties:
\bi
\item The graph is connected, oriented, and cyclically ordered at the vertices.
\item The edges connecting to an incoming pole are all oriented away from the pole, and the edges
corresponding to an outgoing pole are all oriented towards the pole.
\item Each zero connects to cyclically-alternating incoming and outgoing edges, and has a valency of at least four.
\item No edge connects to the same vertex twice, and no edge connects to two poles.
\item Every face of the ribbon graph bounds exactly two poles, one incoming and one outgoing.
\ei
The unique oriented graph $\cG$ associated to the surface $X$ with labelled points $P_i$ and residues $r_i$
is the {\bf Nakamura graph} of the surface.
By assigning the labels $1,2,\ldots, n$ to the vertices of the graph corresponding to the poles $P_1, P_2, \ldots, P_n$, the {\bf pole-labelled Nakamura graph} $\bar{\cG}$ of the surface is also generated from the Giddings-Wolpert differential.

Each face of the graph is bounded by two extended real trajectories of the differential. It is possible to choose local coordinates 
$z$ on each face such that $\omega=dz$ within the face, with $z$ in the range
$0 < \mathrm{Im}(z) < b_j$ for some $b_j$. 
Each face of the graph is holomorphic to a strip $\mathbb{R}\times (0,b_j)$ in the complex plane,
and each strip has a width $b_j$ which is determined by the differential.
The zeroes of the differential lie on the boundaries of the strips.
We can consistently choose complex coordinates on each strip such that the real coordinates of the zeroes on the boundary of the strip match 
the global time coordinates of the zeroes.
The global time coordinates $t_k$ of the zeroes $Q_k$ are unique up to a simultaneous time translation of all the zeroes, $t_k\mapsto t_k+c$. 
This time-translation symmetry can be fixed by putting a constraint on the time coordinates, such as requiring that $\sum_k t_k = 0$.
In this way, each punctured Riemann surface with residues $(\hat{X}, r_i)$ has a unique decomposition into strips via the Giddings-Wolpert differential,
and with the gluing at the strip boundaries determined by the Nakamura graph. We call this the {\bf strip decomposition} of the surface.
The surface $(X, P_i, r_i)$ with labelled points can be recovered from $\hat{X}$ by reintroducing the points at infinity of the strips, corresponding to the poles of the Giddings-Wolpert differential.

\begin{figure}[t]
        \centering
\subfloat[]{\includegraphics[height=0.22\textheight]{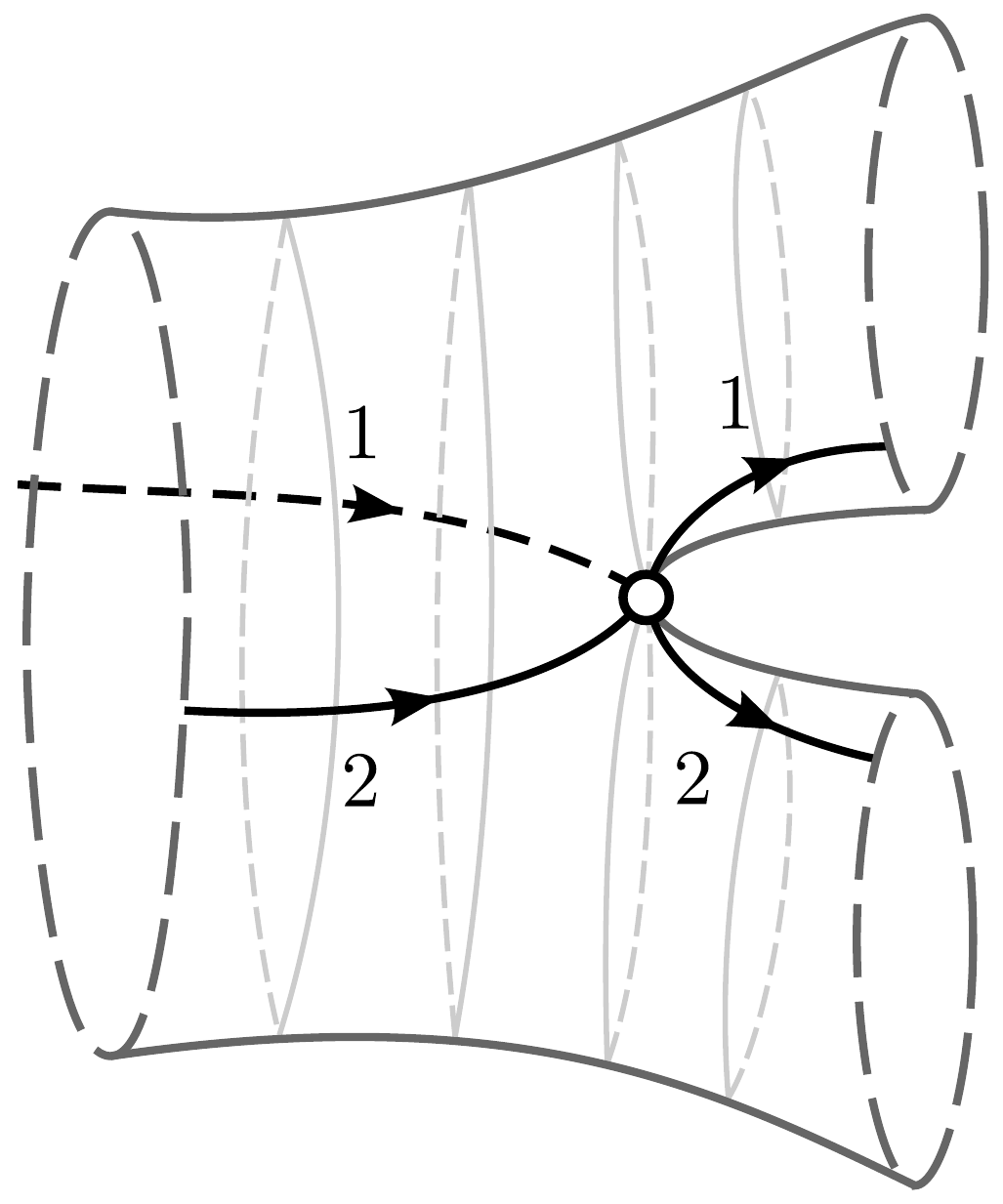} \label{fig:intro-pantsa}}\quad
\subfloat[]{\includegraphics[height=0.22\textheight]{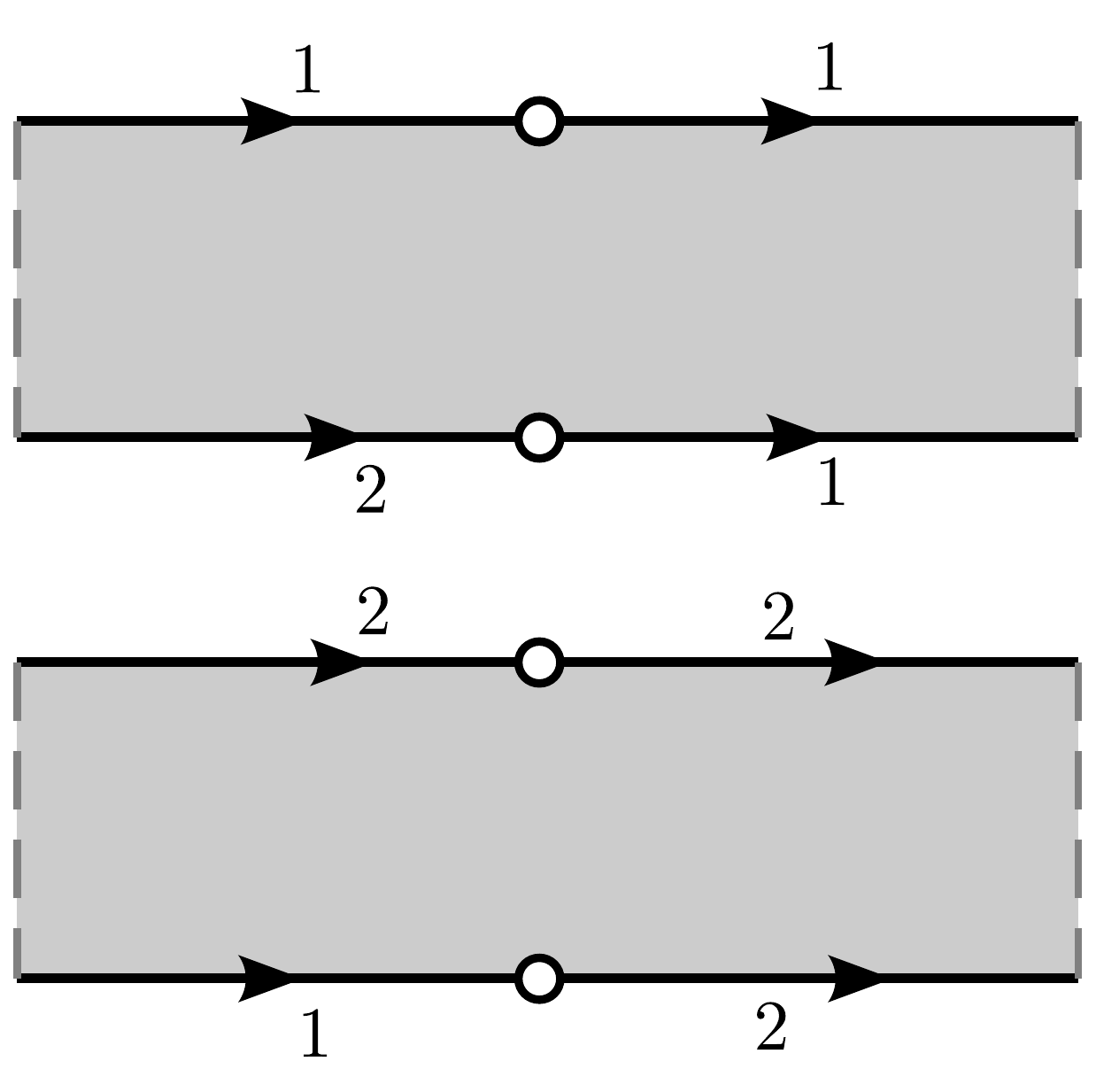} \label{fig:intro-pantsb}}\quad
\subfloat[]{\includegraphics[height=0.22\textheight]{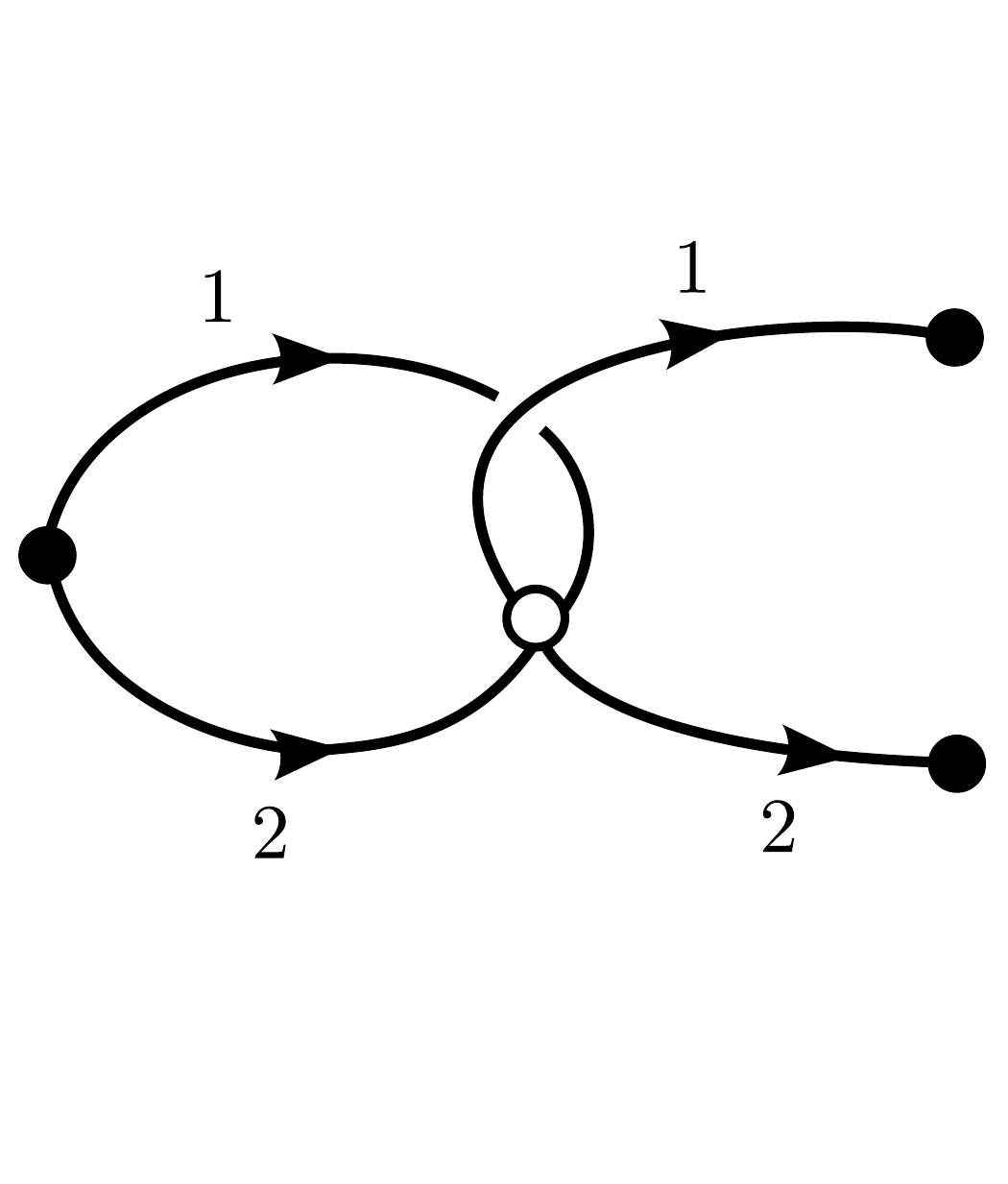} \label{fig:intro-pantsc}}
        \caption{The real trajectories of a Giddings-Wolpert differential on a three-punctured sphere, the associated strip-decomposition, and the Nakamura graph.}\label{fig:intro-pants}
\end{figure}

An example of a Nakamura graph and a strip decomposition of a surface of genus zero with three punctures is given in Figure \ref{fig:intro-pants}. 
In the first part of the figure, the three boundary circles represent the removed points from the surface, taken to infinity, corresponding to the poles of the Giddings-Wolpert differential. 
The white vertices correspond to the zero of the differential. 
We label the graph edges by assigning the same integer to all the edges on the upper boundary of a given strip.
The embedded graph partitions the surface into two strips in the complex plane, which we have drawn in the second figure.
The bounding edges with the same label {\it and the same real coordinate range} are identified.
The third figure shows the Nakamura graph without an embedding into a surface. The black vertices represent the poles of the differential, which are the labelled points of the closed surface or the removed points of the punctured Riemann surface.

Once we have fixed a set of $n$ residues $r_i$, then for each distinct Riemann surface $X$ with $n$ labelled points there is a unique pole-labelled Nakamura graph $\bar\cG$, set of strip widths $b_j$, and set of interaction times $t_k$, up to relabellings of the parameters.
Alternatively, given a Nakamura graph $\bar\cG$ and a consistent set of strip widths $b_j$ and interaction times $t_k$, then we can uniquely reconstruct the associated Riemann surface by constructing and gluing together the holomorphic strips.
For a given graph $\bar\cG$, the set $C(\bar\cG)$ of Riemann surfaces with this graph can be parametrised
by the admissible strip widths $b_j$ and interaction times $t_k$.
One of the main goals of this paper is to specify the set $C(\bar\cG)$ explicitly for any
Nakamura graph, and to show that the collection of all such $C(\bar\cG)$ gives a cell decomposition
of the moduli space $\cM_{g,n}$.

\subsection{Equivalence classes of permutation tuples}

The combinatorics of Nakamura graphs can be described by equivalence classes of permutation tuples in several distinct ways \cite{fgr}.
In this paper, we focus on one description in terms of {\it Hurwitz classes} and {\it slide-equivalence classes}. This description was derived in \cite{fgr} by constructing
branched coverings from the strip decomposition of a surface onto the sphere. 
Here, we review the resulting description.

Consider a Riemann surface $X$ with a strip decomposition $(\cG, b_i, t_j)$.
Let $d$ be the number of strips of the surface, and let $l$ be the number of zeroes
on the surface. It is possible for distinct zeroes of the differential to have equal time coordinates;
let $m\leq l$ be the number of distinct time coordinates of the zeroes.
To each strip of the surface, assign a single label $i\in\{1,2,\ldots, d\}$ to all the edges on the upper
boundary of this strip. This gives a labelling of the edges of the associated Nakamura graph.

We can associate a permutation cycle to each pole and zero of the Giddings-Wolpert differential.
For a pole incident to the (clockwise) cyclically-ordered edges labelled $i_1, i_2\ldots, i_r$, we associate the permutation cycle $(i_1i_2\ldots i_r)$.
For a zero with the clockwise cyclically-ordered {\it incoming} edges $i_1, i_2\ldots, i_s$, we associate the permutation cycle  $(i_1i_2\ldots i_s)$.
This zero will also have the cyclically-ordered outgoing edges $i_1, i_2\ldots, i_s$, with the ingoing edge $i_k$ appearing after the outgoing edge $i_k$ in 
the clockwise ordering.
We can collate the cycles corresponding to incoming poles into a single permutation $\sigma_+\in S_d$, and similarly the outgoing poles into a permutation $\sigma_-\in S_d$. 
To each of the $m$ {\it distinct} time coordinates, we associate a permutation $\sigma_j$ which collates
all the cycles associated to the zeroes with that time coordinate. In this way, we construct a tuple of permutations 
\bea
(\sigma_+, \sigma_1,\ldots, \sigma_m, \sigma_-)
\eea
that describes the labelled Nakamura graph. The product of these $(m+2)$ permutations is the identity permutation.

\begin{figure}[t]
        \centering
\subfloat[]{\includegraphics[width=0.33\textwidth]{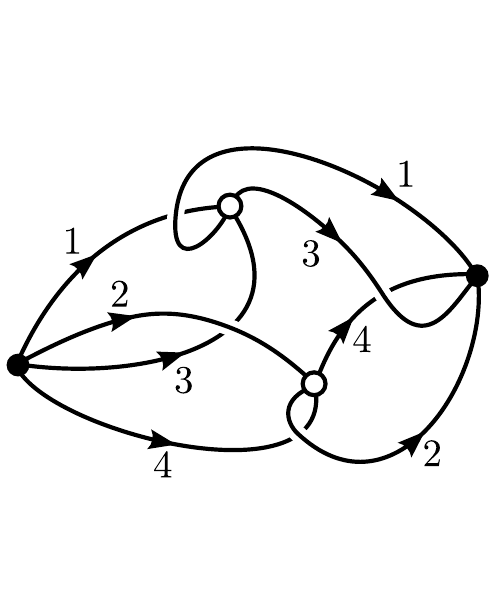}  \label{fig:intro-graphd4}}\qquad
\subfloat[]{\includegraphics[width=0.33\textwidth]{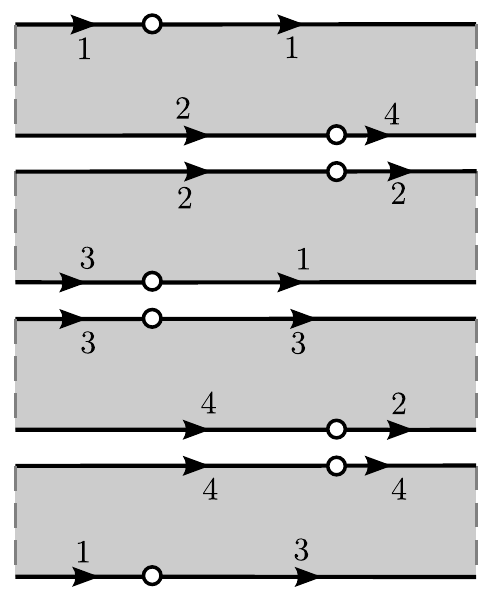}  \label{fig:intro-stripd4}}
\caption{A Nakamura graph and a strip decomposition generated from a genus one surface with two punctures.}
\label{fig:intro-d4}
\end{figure}

An example of a Nakamura graph corresponding to a surface decomposed into four strips is given in Figure \ref{fig:intro-d4}. This graph is described by the tuple
\bea
(\sigma_+, \sigma_1, \sigma_2, \sigma_-) = ((1234),(13),(24), (1234)).
\eea
There is a single incoming pole and a single outgoing pole of this graph, corresponding to the permutations $\sigma_+$ and $\sigma_-$. The two zeroes have distinct time coordinates and correspond to the permutations $\sigma_1$ and $\sigma_2$ respectively.

A Nakamura graph can be uniquely constructed from such a permutation tuple.
However, there can be many different permutation tuples that describe the same Nakamura graph.
There are two equivalence relations on the set of permutation tuples that are required for distinct equivalence classes to correspond to distinct graphs. 
Firstly, the permutation tuples satisfy {\it Hurwitz-equivalence}, which reflects the arbitrary choice of labels
assigned to each edge of the graph. Given a labelling of the edges of the graph, then replacing each edge label $i$ 
with another edge $\gamma(i)$ for some $\gamma\in S_d$ will give the same graph. 
This can be viewed as quotienting out the set of tuples by the automorphism action $\sigma\mapsto \gamma^{-1}\sigma \gamma$ acting simultaneously on 
all the permutations associated to the poles and zeroes.

Secondly, there is the {\it slide-equivalence} relation on the tuples. While a given strip decomposition of a surface determines an ordering of the 
time coordinates of the zeroes, there can be many different surfaces with this graph with different orderings of the time coordinates of the zeroes.
For example, the graph given in Figure \ref{fig:intro-graphd4} could be associated to surfaces with the respective permutation tuple descriptions
\bea
(\sigma_+, \sigma_1, \sigma_2, \sigma_-) &=& ((1234), (13), (24), (1234)), \label{eq:se1} \\
(\sigma_+, \sigma_1,  \sigma_-) &=& ((1234), (13)(24), (1234)), \label{eq:se2} \\
(\sigma_+, \sigma_1, \sigma_2, \sigma_-) &=& ((1234), (24), (13), (1234)), \label{eq:se3}
\eea
depending on the relative time coordinates of the zeroes.
In general, we consider a pair of permutation tuples
\bea
(\sigma_+, \sigma_1, \ldots, \sigma_k, \sigma_{k+1}, \ldots, \sigma_-) \sim (\sigma_+, \sigma_1, \ldots, \sigma_k \sigma_{k+1}, \ldots, \sigma_-)
\eea
to be {slide-equivalent} if the permutations $\sigma_k$ and $\sigma_{k+1}$ are disjoint.
This extends to an equivalence relation on all permutation tuples describing Nakamura graphs, which also commutes with Hurwitz equivalence.
The three tuples given above in \eref{eq:se1}-\eref{eq:se3} are all slide-equivalent.

Within each slide-equivalence class, there exists a unique tuple that minimises the number 
of distinct required permutations, and collates as many cycles into the earliest possible permutations in the tuple as possible. This is called the {\it reduced tuple} of the graph.
In the example above, the reduced tuple is \eref{eq:se2}, as this has fewer permutations than the other slide-equivalent tuples.
By convention, we write reduced tuples with permutations $\tau_j$ associated to the zeroes, instead of $\sigma_j$.

There is a one-to-one correspondence between the combined slide and Hurwitz equivalence classes and the Nakamura graphs.
This correspondence was used in \cite{fgr} to count graphs computationally by counting the Hurwitz-equivalence classes of reduced tuples.

Each graph $\cG$ has an {\it automorphism group} Aut$(\cG)$, which is the set of orientation-preserving mappings of the vertices, edges, and faces of the graph to itself that preserves the graph. These automorphisms necessarily map incoming poles to incoming poles and outgoing poles to outgoing poles. 
For any surface with an embedded labelled graph, the graph automorphisms extend to mappings of the strips to themselves, and so each graph automorphism corresponds to a unique permutation in $S_d$,
and Aut$(\cG)$ can be interpreted as a subgroup of $S_d$.
The automorphism group of a graph coincides with the automorphism group of the graph's reduced tuple, so
Aut$(\cG)$ is the group of permutations $\gamma\in S_d$ that satisfy
\bea
(\gamma^{-1}\sigma_+\gamma, \gamma^{-1}\tau_1\gamma,\ldots, \gamma^{-1}\tau_m\gamma, \gamma^{-1}\sigma_-\gamma) = (\sigma_+, \tau_1,\ldots, \tau_m, \sigma_-).
\eea
%These automorphisms will in general interchange the incoming poles of the graph with themselves, and the outgoing poles of the graph.

\section{Graphs, puncture labellings and moduli space }\label{sec:alt}

There are two distinct notions of the moduli space of Riemann surfaces discussed in \cite{nakamura}, corresponding to the more conventional definition of moduli space 
$\cM_{g,n}$, and a modified version of moduli space $\cM_{g,1[n-1]}$.
The former is the set of equivalence classes of Riemann surfaces of genus $g$ with $n$ labelled
points under biholomorphisms which preserve the labelling of the points.
The latter is the set of equivalence classes under biholomorphisms which can permute the first
$(n-1)$ labelled points.
The more conventional moduli space $\cM_{g,n}$ is much more commonly-used in the literature,
but the modified moduli space $\cM_{g,1[n-1]}$ is simpler to work with computationally,
as its cell decomposition is coarser than that of $\cM_{g,n}$.

There are also two types of graph automorphisms described in \cite{nakamura}: the pole-permuting automorphisms, and the pole-fixing automorphisms.
The automorphism group $\AG$ is the group of bijective mappings of the unlabelled graph $\cG$ to itself which preserves the structure of the graph. 
Graph automorphisms preserve the orientation of the edges, and so map incoming poles to incoming poles and outgoing poles to outgoing poles, but may permute these poles in general.
The automorphism group $\Afix$ is the subgroup of $\AG$ consisting of the automorphisms which fix each pole separately.

Moduli spaces are orbifolds, and any cell in a cell decomposition of an orbifold is homeomorphic to a subset of $\mathbb{R}^k$ modulo a finite group. 
In later sections, we will explicitly show that the automorphism group of a graph corresponds to the orbifold quotienting group of its corresponding cell in moduli space.
The orbifold groups of the cells in $\cM_{g,1[n-1]}$ correspond to the pole-permuting automorphism groups $\AG$, and the orbifold groups of the cells in $\cM_{g,n}$ correspond to the pole-fixing
automorphism groups $\Afix$.
Cells in  $\cM_{g,n}$  are naturally described by graphs with a labelling on the poles,
and cells in $\cM_{g,1[n-1]}$ are described by graphs with no labelling on the poles.
Graphs without pole labellings are described by equivalence classes of permutation tuples, as presented in \cite{fgr} and reviewed in the previous section.

In this section, we introduce a modification of the tuples description of Nakamura graphs that can describe graphs with labelled poles. This extends the approaches to finding the cell decomposition of  $\cM_{g,1[n-1]}$ presented in \cite{nakamura, fgr} and allows us to find the equivalent graphical cell decomposition of the more conventional moduli space $\cM_{g,n}$. 
We also discuss the Euler characteristic of the two types of moduli space, which was used in \cite{nakamura, fgr} as a as highly non-trivial check of the validity of the cell decomposition via Nakamura graphs.

\subsection{Split tuples}\label{sec:splittuples}

%We are free to choose any values for the residues $r_i$ of the poles, so in the following
%we set exactly one of the residues $r_n$ to be positive, and the remaining residues $r_1,\ldots, r_{n-1}$ to be negative. The corresponding Nakamura graphs will have one incoming pole and $(n-1)$ outgoing poles.
%When considering Nakamura graphs corresponding to cells in $\cM_{g,1[n-1]}$, 
%we must also set the negative residues to be equal: $r_1=r_2=\ldots=r_{n-1}$.

Let $\cG$ be a graph with edges labelled from $1$ to $d$ with a single incoming pole and $(n-1)$ outgoing poles.
This graph corresponds to a slide-equivalence class of permutation tuples of the form 
$(\sigma_+,\sigma_1,\ldots, \sigma_m, \sigma_-)$, where $\sigma_+$ is a single cycle and $\sigma_-$ is composed of $(n-1)$ cycles.
Let $\bar\cG$ be the same graph with the labelling $\{1,2,\ldots, n-1\}$ assigned
to the outgoing poles, and the label $n$ assigned to the incoming pole.
Each cycle of $\sigma_-$ corresponds to an outgoing pole of the graph, labelled from 1 to $(n-1)$.
We write $\sigma_-^{(i)}$ to denote the cycle corresponding to the $i$th pole of the labelled
graph.
A {\bf split tuple} is a tuple arising from a Nakamura graph in which $\sigma_-$ is replaced
by $(n-1)$ disjoint ordered single cycles corresponding to the outgoing poles of the graph:
\bea
(\sigma_+, \sigma_1, \sigma_2, \ldots, \sigma_m;\ \sigma_-^{(1)},\sigma_-^{(2)},\ldots, \sigma_-^{(n-1)} ).
\eea
If the Hurwitz class is a reduced tuple of a labelled graph $\bar\cG$, then we call its split tuple a {\bf reduced split tuple}, and replace the $\sigma_k$ with $\tau_k$:
\bea
(\sigma_+, \tau_1, \tau_2, \ldots, \tau_m;\ \sigma_-^{(1)},\sigma_-^{(2)},\ldots, \sigma_-^{(n-1)} ).
\eea
In this subsection, we will constrain ourselves to considering the reduced split tuples of graphs with labelled poles and labelled edges.

A general element $\gamma\in S_d$ acts on a tuple by relabelling each individual cycle in the permutations 
\bea
(j_1 j_2 \ldots j_p) \overset{\gamma}{\longrightarrow} (\gamma(j_1)\gamma(j_2) \ldots \gamma(j_p)). \label{eq:cyclewise}
\eea
This relabelling corresponds to the action by conjugation on the permutations in the tuple $\sigma_+ \mapsto \gamma^{-1}\sigma_+\gamma$, $\tau_k\mapsto \gamma^{-1}\tau_k\gamma$, $\sigma_-^{(i)} \mapsto \gamma^{-1}\sigma_-^{(i)}\gamma$. (For a cycle $(j)$ of length 1, the action of conjugation by $\gamma$ is defined to be $(j) \mapsto (\gamma(j))$.)
The group of permutations that preserve the reduced tuple $(\sigma_+, \tau_1, \ldots, \tau_m, \sigma_-)$ under this action is isomorphic to the group of automorphisms of the unlabelled graph $\AG$.
An automorphism $a\in \AG$ will preserve $\sigma_-$ under conjugation, but will not generally preserve each $\sigma_-^{(i)}$,
and may interchange them.
This means that the action of $a$ on a given single-cycle $\sigma_-^{(i)}$ in $\sigma_-$ will produce another single-cycle $\sigma_-^{(i')}$ in $\sigma_-$,
and so we can read off an element $\kappa_a\in S_{n-1}$ such that the action of the automorphism is
\bea
\sigma_-^{(i)} \overset{a}{\longrightarrow} \sigma_-^{(\kappa_a(i))}.
\eea
The mapping from an automorphism $a\in S_d$ to a pole-permutation $\kappa_a\in S_{n-1}$ is a homomorphism
$\phi: \AG \to S_{n-1}$.
The kernel of this homomorphism is $\Afix$, the group of automorphisms which fix each pole.
By the isomorphism theorem, the group $H:=\AG/\Afix$ is isomorphic to a subgroup of $S_{n-1}$.
This is the group of all $\kappa_a \in S_{n-1}$ arising from the automorphisms $a\in \AG$. Two permutations in this group $\kappa_a$, $\kappa_b$ arising from the distinct automorphisms $a, b \in \AG$ are identical 
if there is some pole-fixing automorphism $a_{\mathrm{fix}}\in \Afix$
with $a= b \circ a_{\mathrm{fix}}$.

\begin{figure}[t]
\begin{center}
\includegraphics[width=0.9\textwidth]{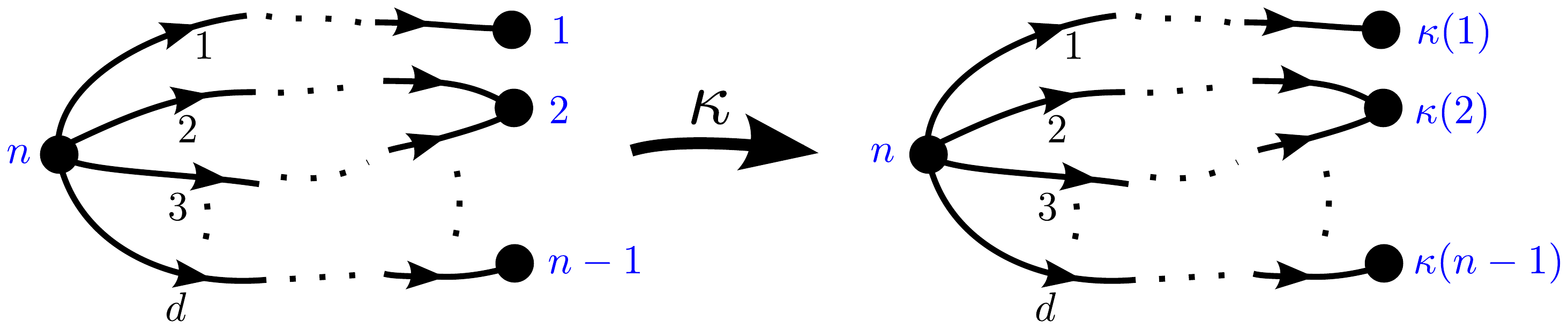} \qquad
\caption{An element $\kappa \in S_{n-1}$ relabels the outgoing poles of a labelled graph $\bar{\cG}$.}
\label{fig:intro-kappapoles}
\end{center}
\end{figure}

A relabelling of the outgoing poles of a graph can be described by a general element $\kappa\in S_{n-1}$ acting on the split tuple by rearranging the $(n-1)$ single cycles
\bea
&(\sigma_+, \tau_1, \tau_2, \ldots, \tau_m; & \sigma_-^{(1)},\sigma_-^{(2)},\ldots, \sigma_-^{(n-1)} ) \label{eq:tuple1} \\
 \overset{\kappa}{\longrightarrow}
 &(\sigma_+, \tau_1, \tau_2, \ldots, \tau_m; & \sigma_-^{(\kappa(1))},\sigma_-^{(\kappa(2))},\ldots, \sigma_-^{(\kappa(n-1))} ).\label{eq:tuple2}
\eea
This action is demonstrated on a general graph in Figure \eref{fig:intro-kappapoles}.
The arrangements of the split cycles given in \eref{eq:tuple1} and \eref{eq:tuple2} correspond to the same graph if there is some relabelling of the graph edges $\gamma\in S_d$ that maps one to the other.
Such a relabelling must preserve $\sigma_+$ and each $\tau_k$ separately, and hence must be an automorphism
$\gamma \in \AG \subset S_d$.
The action of an automorphism $\gamma$ on the $(n-1)$ split-cycles is described by some $\kappa_\gamma \in S_{n-1}$.
We see that the tuples \eref{eq:tuple1} and \eref{eq:tuple2} describe the same graph if and only if
$\kappa$ corresponds to some automorphism, i.e. $\kappa= \kappa_\gamma$ for some $\gamma \in \AG$.
This is precisely the statement that $\kappa$ is in the image of $\AG$ under the homomorphism $\phi$, i.e. $\kappa \in H$.
We conclude that the distinct split tuples associated to a reduced tuple correspond to the distinct {\it cosets} of $H$ in $S_{n-1}$.
For each graph $\cG$ without pole labellings, there are $(n-1)!|\Afix|/|\AG|$ distinct pole-labelled graphs $\{\bar\cG\}$.

\begin{figure}[t]
        \centering
\subfloat[]{\includegraphics[height=0.13\textheight]{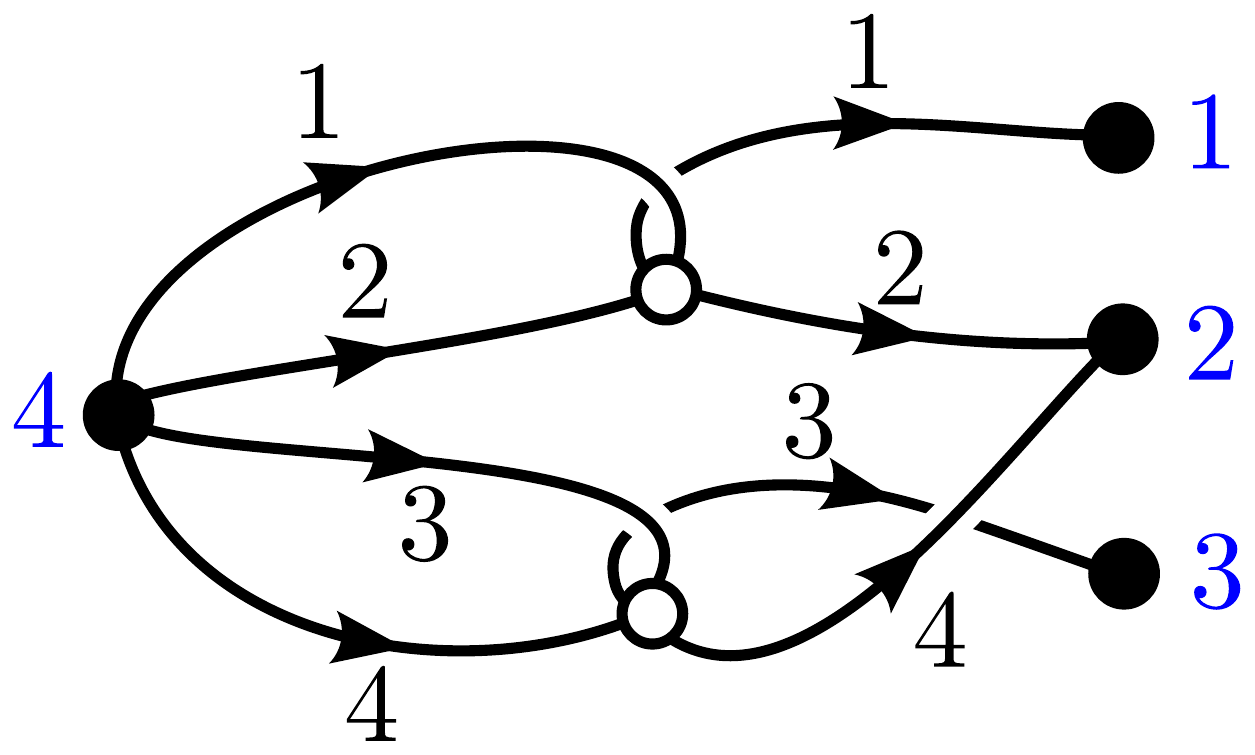}  \label{fig:hurwitz-polesa}}\hspace{0mm}
\subfloat[]{\includegraphics[height=0.13\textheight]{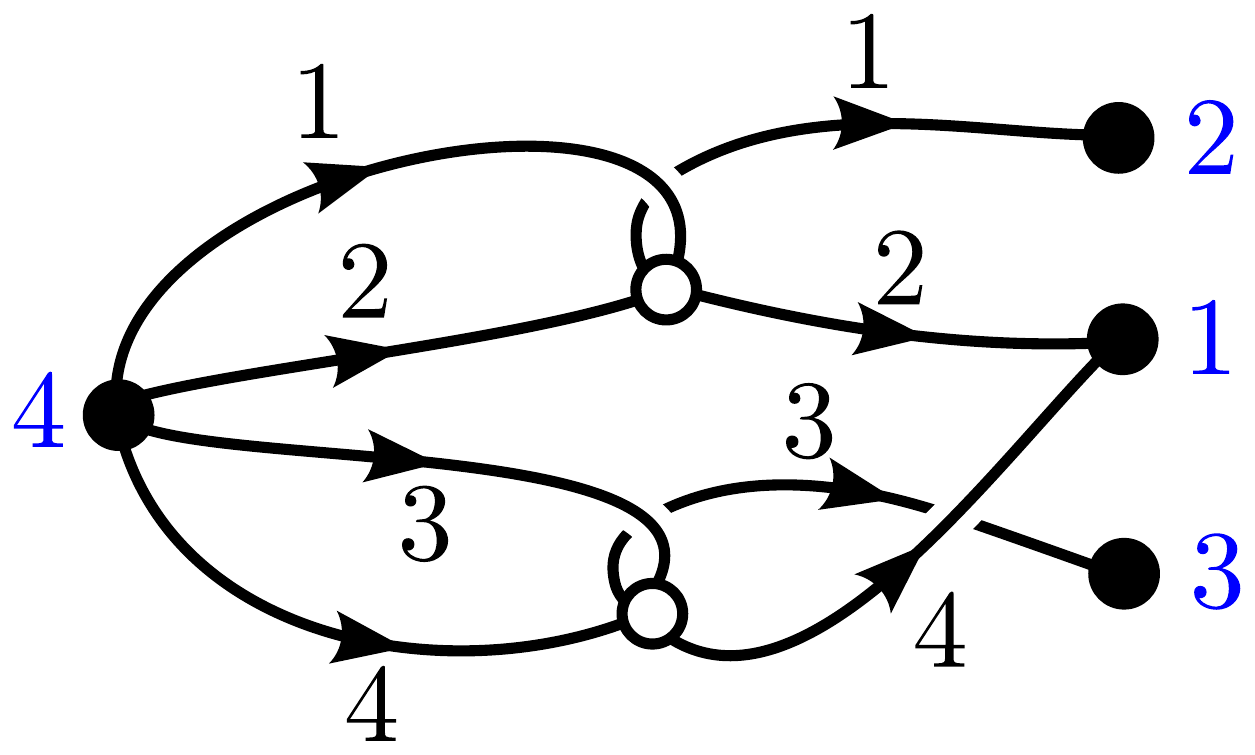}  \label{fig:hurwitz-polesb}}\hspace{0mm}\subfloat[]{\includegraphics[height=0.13\textheight]{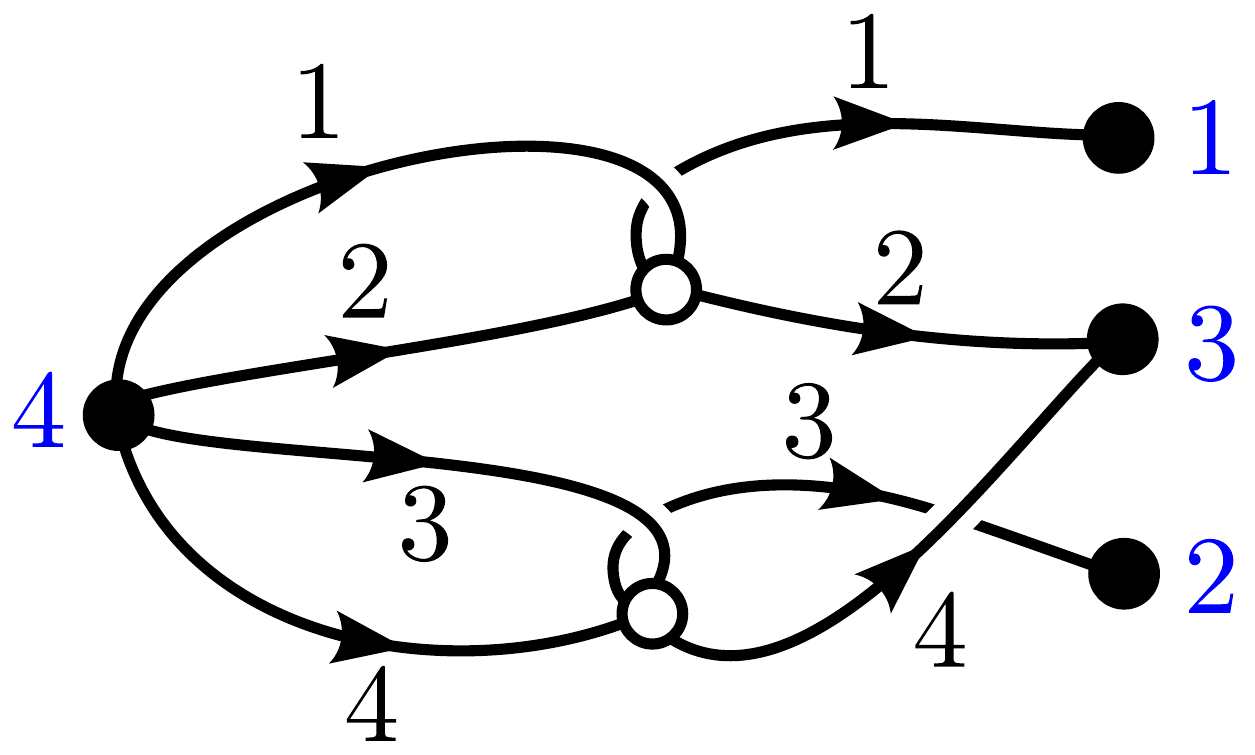}  \label{fig:hurwitz-polesc}}
        \caption{A Nakamura graph with edge labellings and three inequivalent pole labellings.}\label{fig:hurwitz-poles}
\end{figure}

As an example, we consider the graph with $n=4$ poles described by the unlabelled tuple
\bea
(\sigma_+, \tau, \sigma_-) = ( (1234), (12)(34), (1)(24)(3) ). \label{eq:unlabelledtuple}
\eea
We can split $\sigma_-$ into the constituent cycles $\sigma_-^{(1)}, \sigma_-^{(2)},\sigma_-^{(3)}$:
\bea
(\sigma_-^{(1)}, \sigma_-^{(2)},\sigma_-^{(3)})  = ( (1), (24), (3) ). \label{eq:splitcycles}
\eea
The distinct pole labellings correspond to the unique ways to order the permutations $\sigma_-^{(1)}$, $\sigma_-^{(2)}$, $\sigma_-^{(3)}$ in the tuple, up to edge-relabellings. 
We can derive the pole labellings from the cosets construction as follows.
The automorphism group of the unlabelled tuple in Equation \eref{eq:unlabelledtuple} is isomorphic to $\mathbb{Z}_2$, generated
by the action by conjugation of the permutation $\gamma:=(13)(24)$ on each element of the tuple.
This automorphism maps the split-cycles as follows:
\bea
\sigma_-^{(1)}  & \overset{\gamma}{\longrightarrow}& \sigma_-^{(3)}, \ret
\sigma_-^{(2)} & \overset{\gamma}{\longrightarrow} &\sigma_-^{(2)}, \ret
\sigma_-^{(3)} & \overset{\gamma}{\longrightarrow}& \sigma_-^{(1)},
\eea
and so defines an element $\kappa_\gamma:=(13)\in S_3 = S_{n-1}$. The cosets of $H:=\cor{(13)}$ in $S_3$ correspond
to the distinct ways of rearranging the outgoing poles of the tuple. The cosets are $H$, $(12)H$ and $(23)H$, and we can choose the representative elements of these cosets $\{ (), (12), (23)\}$.
Acting with these representative elements on the split-cycles in Equation \eref{eq:splitcycles} gives the three inequivalent split tuples associated to the pole-labelled graphs, shown in Figure \eref{fig:hurwitz-poles}:
\bea
(\sigma_+, \tau ;\ \sigma_-^{(1)}, \sigma_-^{(2)},\sigma_-^{(3)} ) &=&  ( (1234), (12)(34); \ (1), (24), (3)  ), \label{eq:firstone} \\
(\sigma_+, \tau ;\ \sigma_-^{(2)}, \sigma_-^{(1)},\sigma_-^{(3)} ) &=&  ( (1234), (12)(34); \ (24), (1), (3)  ),\\
(\sigma_+, \tau ;\ \sigma_-^{(1)}, \sigma_-^{(3)},\sigma_-^{(2)} ) &=&  ( (1234), (12)(34); \ (1), (3), (24)  ).
\eea
As an example of a split tuple which is equivalent to one of these tuples, we
can see that acting on the poles in Equation \eref{eq:firstone} with the permutation $(13)\in H$ gives the tuple
\bea
(\sigma_+, \tau; \ \sigma_-^{(3)}, \sigma_-^{(2)},\sigma_-^{(1)} ) &=&  ( (1234), (12)(34);\ (3), (24), (1) ),
\eea
which describes the same graph as the tuple in Equation \eref{eq:firstone} as the two tuples can be related by the relabelling automorphism $\gamma=(13)(24)$.

\subsection{Euler characteristics}

The orbifold Euler characteristic was used in \cite{nakamura, fgr} to verify the cell decomposition of $\cM_{g,1[n-1]}$.
If $\{C\}$ is a cell decomposition of an orbifold such that the orbifold group at every point in a given cell $C$ is constant, then the orbifold Euler characteristic is \cite{thurston}
\bea
\chi(\{C\}) = \sum_{C} (-)^{\mathrm{dim}(C)}\frac{1}{|A(C)|},
\eea
where $A(C)$ is the orbifold group at any point in the cell $C$.
The Nakamura graph cell decomposition of $\cM_{g,1[n-1]}$ allows us to write this formula in terms of a sum over graphs and their parameters;
\bea
\chi(\cM_{g,1[n-1]}) =  \sum_{\cG} (-)^{d+l-n}\frac{1}{|\mathrm{Aut}(\cG)|}, \label{eq:coarse}
\eea
where $d$ is the number of faces, $l$ the number of zeroes, and $n$ the number of poles of a graph.
The analogous formula for Euler characteristic of the moduli space $\cM_{g,n}$ is
\bea
\chi(\cM_{g,n}) =  \sum_{\bar{\cG}} (-)^{d+l-n}\frac{1}{|\mathrm{Aut}(\bar{\cG})|}, \label{eq:fine}
\eea
where the sum runs over the distinct pole-labelled graphs $\{\bar\cG\}$.

In the previous section, we showed that each graph $\cG$ without pole labellings
corresponds to $(n-1)!|\Afix|/|\AG|$ graphs with pole labellings, corresponding
to the cosets of $H=\AG/\Afix$ in $S_{n-1}$.
The automorphism group $\mathrm{Aut}(\bar{\cG})$ of any labelling of a graph $\cG$ is isomorphic
to $\Afix$, and so we can see that
\bea
\sum_{\bar{\cG}} (-)^{d+l-n}\frac{1}{|\mathrm{Aut}(\bar{\cG})|} &=& \sum_{\cG} (-)^{d+l-n}\frac{(n-1)!}{|\Afix|}  \frac{|\Afix|}{|\AG|} \ret
&=& (n-1)! \sum_{\cG}(-)^{d+l-n} \frac{1}{|\AG|},
\eea
and so the Euler characteristics of the two different moduli spaces differ by a factor of $(n-1)!$:
\bea
\chi(\cM_{g,n}) = (n-1)! \chi(\cM_{g,1[n-1]}).
\eea
This relation was used to compare the Euler characteristic of the Nakamura graph cell decomposition of $\cM_{g,1[n-1]}$ with the exact expressions of $\chi(\cM_{g,n})$ given in \cite{hz}.
The Euler characteristics of $\chi(\cM_{g,n})$ were found to match for all moduli spaces with $(2g-2+n)\leq 7$.

\section{Cell decompositions}\label{sec:celldecomp}

In this section, we explicitly show how to construct the set of points in moduli space $\cM_{g,n}$ associated to a pole-labelled graph $\bar\cG$ by using its reduced split tuple.
We find that this set is in one-to-one correspondence with a subset $\cB(\bar\cG)$ in $\mathbb{R}^{d+l}$, modulo the action of the pole-fixing graph automorphism group $\mathrm{Aut}(\bar\cG)$.
The subset $\cB(\bar\cG)$ is a  $ ( d + l - n ) $ dimensional  {\it convex polytope}, which is an intersection of finitely many real hyperplanes and half-spaces given by a system of linear equalities and inequalities \cite{Ziegler}.  

We explain how to find the boundaries and incidences of the sets  $\cB(\bar\cG)/\mathrm{Aut}(\bar\cG)$  from the split tuples.
This confirms the claim of \cite{nakamura} that the set of distinct Nakamura graphs give a valid cell decomposition of moduli space.
We also discuss the generalisation of the cell decomposition to the moduli space $\cM_{g,1[n-1]}$,
and show that a cell $C(\cG)$ corresponding to a graph without pole labellings is described
by a set homeomorphic to $\mathbb{R}^{d+l-n}$ modulo the pole-permuting automorphisms $\AG$.

As extended examples, we give the graph decompositions of the low-dimensional moduli spaces
$\cM_{0,4}$ and $\cM_{1,2}$, and show that the cell decomposition of $\cM_{0,4}$ matches
the description of the space known by considering M\"obius maps on the sphere.
We briefly describe how to construct the moduli space $\cM_{0,1[3]}$ from the quotienting
of $\cM_{0,4}$. We conclude this section by discussing the generalisation of this cell decomposition to Teichm\"uller space.

\subsection{Cells in moduli space}\label{sec:cells}

Recall that any cell in a cell decomposition of an orbifold is homeomorphic to a ball modulo a finite group. The aim of this subsection is to show that, from the reduced split tuple of any labelled Nakamura graph $\bar\cG$, we can construct a convex polyhedron $\cB(\bar\cG)$ on which $\mathrm{Aut}(\bar\cG)$ acts in such a way that $\cB(\bar\cG)/\mathrm{Aut}(\bar\cG)=C(\bar{\cG})$.

Consider the moduli space of inequivalent Riemann surfaces of genus $g$ with $n$ labelled points $\cM_{g,n}$, and choose a set of negative reals $r_1,\ldots, r_{n-1}$ with the positive real $r_n=-\sum_i r_i$.
Let $C(\bar{\cG})$ be the collection of points in $\cM_{g,n}$ with the same labelled Nakamura graph $\bar{\cG}$ associated to the reduced split tuple
\bea
(\sigma_+, \tau_1, \tau_2, \ldots, \tau_m;\ \sigma_-^{(1)},\sigma_-^{(2)},\ldots, \sigma_-^{(n-1)} ). \label{eq:polefixedtuple}
\eea
A Riemann surface with this Nakamura graph is specified by a set of $d$ strip widths and $l$ interaction times.
Label the widths of the strips $b_1, b_2, \ldots, b_d$  and the interaction times of the zeroes $t_1, t_2, \ldots, t_l$.
The cell $C(\bar{\cG})$ is parametrised by a subspace of $\mathbb{R}^{d+l}$. 
The $(d+l)$ variables parametrising the cell are subject to some linear constraints determined by the structure of the graph and our choice of time-symmetry fixing:
\begin{itemize}
\item The strip widths must be consistent with the residues at the poles. 
Each single cycle $\sigma_-^{(i)} = (j_1 j_2 \ldots j_k)$ corresponds to
the $i$th pole of the graph, which connects to $k$ strips with widths $b_{j_1}, b_{j_2}, \ldots, b_{j_k}$. This gives $(n-1)$ independent constraints of the form
\bea
b_{j_1}+ b_{j_2}+ \ldots + b_{j_k} = |r_i|. \label{eq:residuesum}
\eea
(There is also a constraint that the total sum of the widths of all strips must correspond to $\sum |r_i|$, but this constraint is derived from the above $(n-1)$ constraints.)

\item The time-translation symmetry of the zeroes is fixed by requiring that the sum over
all the interaction times is zero,
\bea
t_1 + t_2 + \ldots + t_l = 0. \label{eq:timesum}
\eea
\end{itemize}

Each of these $n$ constraints is of the form $\sum q_j = C$ for some continuous parameters $q_j$ and some constant $C$,
which are the equations of a set of hyperplanes in $\mathbb{R}^{d+l}$. 
As well as the above hyperplane constraints, there are some `half-space' constraints on the variables, which are of the form 
$\sum q_j > C$. 
Each of these constraints partitions $\mathbb{R}^{d+l}$ into two subsets by a hyperplane, and so does not lower the dimension of the space. The independent half-space constraints are formulated 
from the reduced split tuple as follows:
\begin{itemize}
\item Each strip-width $b_j$ must be positive. However, if a label $j \in \{1,\ldots d\}$ corresponds to some 1-cycle in $\sigma_-$, i.e. $\sigma_-^{(i)} = (j)$, 
then the constraint $b_j = |r_i|$ automatically implies that $b_j >0$. (Geometrically, the hyperplanes $b_j=0$ and $b_j =|r_i|$ are parallel.)
The independent half-plane constraints on the strip widths are
\bea
b_j>0, \qquad j\in\{1,2,\ldots, d\} \mathrm{\ and\ } \sigma_-(j)\neq j. \label{eq:widthineq}
\eea

\item The interaction times of the zeroes must respect the ordering of the associated cycles of the reduced tuple. 
Recall that a given $\tau_k$ in the reduced tuple consists of multiple disjoint non-trivial single cycles of the form
\bea
\tau_k = \sigma_{p_1}\sigma_{p_2}\ldots\sigma_{p_{a}},
\eea
where $\{p_1, p_2, \ldots, p_a\}\subset \{1,2,\ldots, l\}$.
Each $\sigma_{p_j}$ corresponds to a zero of the Giddings-Wolpert differential on the surface with interaction time $t_{p_j}$. It follows explicitly from the construction of the reduced tuple that every cycle in $\tau_{k+1}$ must appear at a later interaction
time than every cycle in $\tau_k$. If $\tau_{k+1}$ has the decomposition
\bea
\tau_{k+1} = \sigma_{q_1}\sigma_{q_2}\ldots\sigma_{q_{b}},
\eea
for some  $\{q_1, q_2, \ldots, q_b\}\subset \{1,2,\ldots, l\}\backslash \{p_1, p_2, \ldots, p_a\}$,
then we have a half-plane constraint
\bea
t_{p_\alpha}< t_{q_\beta} \label{eq:timeineq}
\eea
for each pair $(p_\alpha, q_\beta)$ with $p_\alpha\in\{p_1, p_2, \ldots, p_a\}$
and $q_\beta \in  \{q_1, q_2, \ldots, q_b\}$.
The collection of all such constraints are necessary and sufficient to guarantee that any configuration of the interaction times is consistent with the chosen Nakamura graph.
If $\tau_1, \tau_2, \ldots, \tau_m$ each have $c_1, c_2, \ldots c_m$ constituent non-trivial
cycles respectively, where $c_1+c_2+\ldots + c_m=l$, then there are
\bea
c_1c_2 + c_2c_3+\ldots + c_{m-1}c_m
\eea
inequalities imposed on the interaction times.

A figure demonstrating the time-ordering inequalities associated to a permutation tuple is given in Figure \ref{fig:cells-torder}. In this example, the single-cycle interaction permutations $\sigma_k$ are collated into the reduced tuple permutations $\tau_k$ with $\tau_1=\sigma_1\sigma_2$, $\tau_2=\sigma_3\sigma_4\sigma_5$, $\tau_3=\sigma_6$, and $\tau_4 = \sigma_7 \sigma_8$. Each line connecting the consecutive $\sigma_i$ corresponds to an inequality on the time coordinates of the respective interaction points.
\begin{figure}[t]
\begin{center}
\includegraphics[height=0.2\textheight]{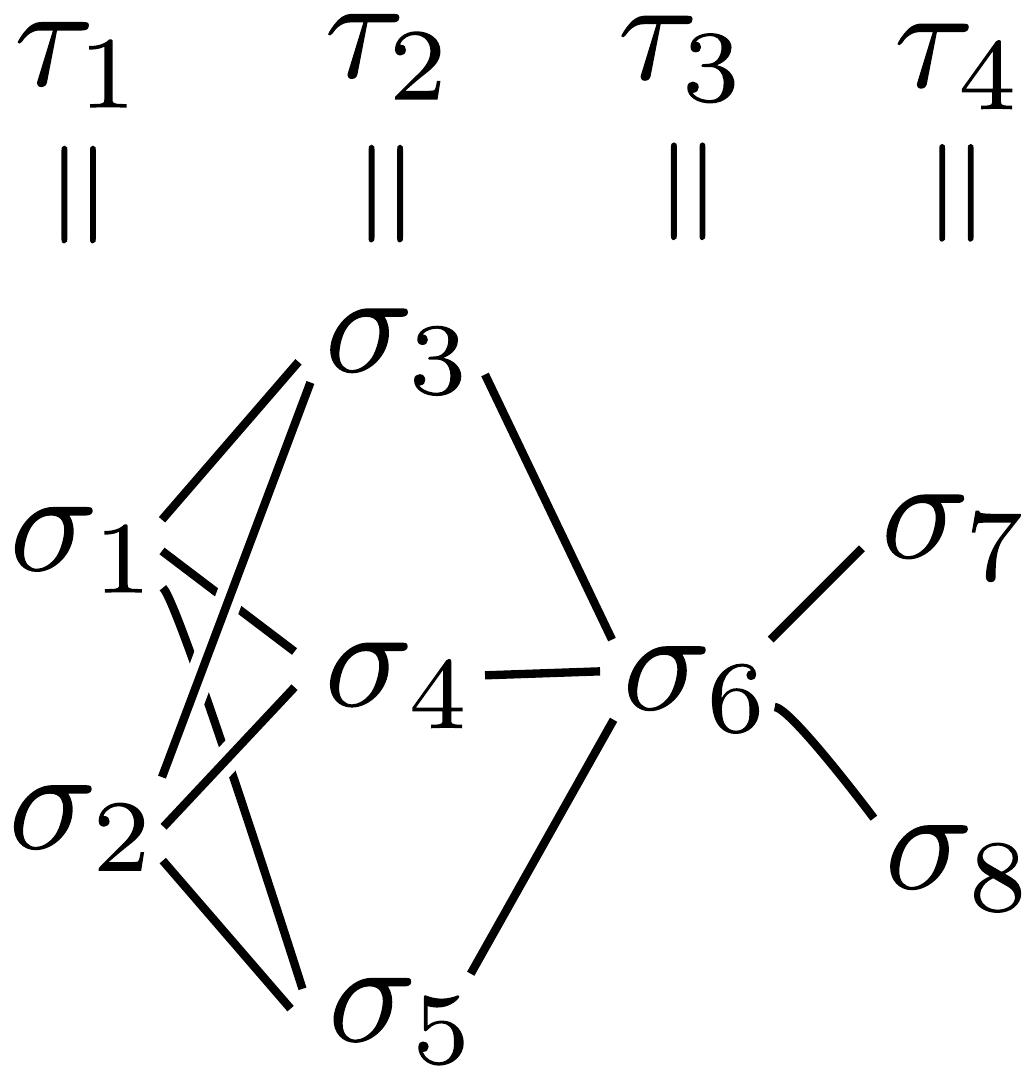} \qquad
\caption{Each line corresponds to a time-ordering inequality.}
\label{fig:cells-torder}
\end{center}
\end{figure}

\end{itemize}

Each of these hyperplane and half-space constraints defines a convex subspace of $\mathbb{R}^{d+l}$.
The intersection of convex subspaces is convex, and so these constraints define a convex subspace of $\mathbb{R}^{d+l}$, which is a polytope.
The $n$ hyperplane constraints define subspaces of $\mathbb{R}^{d+l}$ of codimension one, and the half-space constraints
define subsets of codimension zero, hence these constraints define a convex $(d+l-n)$-dimensional subset of $\mathbb{R}^{d+l}$,
which is homeomorphic to $\mathbb{R}^{d+l-n}$.
We denote this convex polytope by $\cB(\bar\cG)$.

The group $\mathrm{Aut}(\bar\cG)=\Afix$ of pole-fixing automorphisms of the tuple \eref{eq:polefixedtuple} has a natural action on the parameters of $\cB(\bar\cG)$.
An element $a_{\mathrm{fix}}\in \mathrm{Aut}(\bar\cG)\subset S_d$ acts on the strip widths by the relabelling
\bea
b_j \to b_{a_{\mathrm{fix}}(j)}.
\eea
It also acts on the interaction vertices by permutation: recall that the action by conjugation of an automorphism $\afix$ will 
fix any $\tau_k$, but may permute around the constituent cycles $\sigma_{p_1}, \ldots, \sigma_{p_a}$ in $\tau_k$. 
There is therefore some permutation $\kappa$ acting on the integers $\{p_1, p_2, \ldots, p_a\}$ corresponding to the automorphism $\afix$. The action of $\afix$ on the interaction times corresponds to this permutation: 
\bea
t_{p_\alpha} \to t_{\kappa({p_\alpha})}.
\eea

We can see that the defining constraints of $\cB(\bar\cG)$ are preserved under these actions. Pole-fixing automorphisms do not
modify the  split-cycles $\sigma_-^{(i)}$, so the strip-width equations \eref{eq:residuesum}
and inequalities \eref{eq:widthineq} are all preserved.
The interaction times are all permuted into each other, so the overall sum \eref{eq:timesum} is preserved.
Also, the interaction vertices are only permuted within each $\tau_k$ separately: cycles within $\tau_k$ are mapped
to cycles within $\tau_k$. This means that the associated time-ordering inequalities of the form \eref{eq:timeineq} are mapped into each other, and so these constraints are preserved. 
This is enough to conclude that pole-fixing automorphisms map $\cB(\bar\cG)$ into itself.

Finally, we can show that points in $\cB(\bar\cG)$ related by a pole-fixing automorphism
correspond to strip decompositions of Riemann surfaces related by a biholomorphism.
Let $X$ and $\tilde{X}$ be Riemann surfaces with the same labelled graph $\bar\cG$ and 
the respective sets of strip parameters $(b_j, t_k)$ and $(\tilde{b}_j, \tilde{t}_k)$.
If there exists a biholomorphism $f:X\to \tilde{X}$, then the pull-back of the Giddings-Wolpert 
differential from the surface $\tilde{X}$ satisfies the required properties of a Giddings-Wolpert differential on the surface $X$, and so is the Giddings-Wolpert differential of $X$ by uniqueness. This implies that the biholomorphism $f$ preserves 
the strip decomposition of the surface, and so restricts to an automorphism of the graph $a_f\in\mathrm{Aut}(\bar\cG)$ on the strip boundaries, and the action of this automorphism
on the point in $\cB(\bar\cG)$ is $a_f:(b_j,t_k)\mapsto(\tilde{b}_j,\tilde{t}_k)$.
Conversely, if $(b_j, t_k)$ and $(\tilde{b}_j, \tilde{t}_k)$ are points in $\cB(\bar\cG)$
related by a graph automorphism $a\in\mathrm{Aut}(\bar{\cG})$ which maps $b_j\mapsto\tilde{b}_j$ and $t_k\mapsto\tilde{t}_k$, then $\tilde{b}_j=b_{a(j)}$ and $\tilde{t}_k=t_{\alpha(k)}$
for some $\alpha\in S_l$. The strip decompositions of the Riemann surfaces constructed from the parameters $(b_j, t_k)$ and $(\tilde{b}_j, \tilde{t}_k)$ are related by a biholomorphism mapping the strip with upper edges labelled $j$ to the strip with upper edges labelled $a(j)$.
This is enough to conclude the following:

{\bf Theorem: }
Given a Nakamura graph $\bar{\cG}$ with genus $g$, $d$ faces, $l$ interaction points (zeroes), one ingoing pole, and $(n-1)$ outgoing poles labelled $1,2,\ldots (n-1)$ respectively, then there is a convex polytope $\cB(\bar{\cG}) \subset \mathbb{R}^{d+l}$ of dimension $(d+l-n)$ which parametrises the possible strip widths and interaction times of the graph. The group $\mathrm{Aut}(\bar\cG)$ of pole-fixing automorphisms of $\bar\cG$ is a subgroup of the isometries of $\cB(\bar\cG)$, and there is a one-to-one correspondence between the quotient space $\cB(\bar\cG)/\mathrm{Aut}(\bar\cG)$ and the set $C(\bar{\cG})$ of Riemann surfaces with the Nakamura graph $\bar{\cG}$.

We can extend the above description to cells in the modified moduli space $\cM_{g,1[n-1]}$. 
First, we must set the outgoing residues of the $n$ poles to be equal, $r_1=\ldots=r_{n-1}=r$, for some negative real $r$.
For an unlabelled graph $\cG$ with associated reduced tuple
\bea
(\sigma_+, \tau_1,\ldots, \tau_m, \sigma_-),
\eea
the parameter space $\cB(\cG)$ is defined entirely similarly to that of a pole-labelled graph.
Each cycle in $(j_1\ldots j_k)\in\sigma_-$ determines a strip-width constraint
\bea
b_{j_1}+ b_{j_2}+ \ldots + b_{j_k} = |r|. \label{eq:residuesum2}
\eea
The pole-permuting automorphisms $\AG$ interchange the cycles of $\sigma_-$, and so permutes
the set of $(n-1)$ constraints of the form \eref{eq:residuesum2} into each other: this is consistent when the outgoing pole residues are equal.
The quotient space $C(\cG) = \cB(\cG)/\AG$ is therefore well-defined, and the arguments above for the one-to-one correspondence between strip decompositions of Riemann surfaces and the parameter space still hold for pole-permuting automorphisms.
The cells in $\cM_{g,1[n-1]}$ corresponding to graphs without pole labellings can be derived from the quotienting of cells in $\cM_{g,n}$ corresponding to pole-labelled graphs via the group $\AG/\Afix$.

%$
\subsection{Boundaries of cells}\label{sec:bdy}

In terms of the interaction times $t_k$ and the strip widths $b_j$, 
the boundaries of a cell $C(\bar\cG)$ correspond to the limits of the half-spaces of the form $t_p<t_q$ and $b_j>0$. The first of these corresponds to the collapsing of an internal edge, which merges two interaction points together, and the second corresponds to the collapsing of one of the strips of the graph. In the following, we outline how to determine the boundaries of the cells from both the strips description and the tuples description.

Recall that each pole-labelled Nakamura graph is described by an equivalence class of split tuples,
in which tuples are equivalent when they are related by conjugacy or slide-equivalence.
Every point in the cell of the graph has an associated conjugacy equivalence class of split tuples,
generated from the branched covering of the surface onto the sphere and the labelling of the poles.
Zeroes of the Nakamura graph with the same time coordinate will correspond to cycles in the same permutation.
This conjugacy equivalence class will not be an equivalence class of reduced split tuples in general, but will be slide-equivalent to the reduced split tuple of the graph.
The reduced tuple comes from tuning as many of the time-coordinates of the zeroes to be coincident and as early as possible.
We can generate an {\bf expanded split tuple} by tuning the time-coordinates such that every zero of the graph has a distinct time coordinate.

\begin{figure}[t]
\centering
\includegraphics[width=0.9\textwidth]{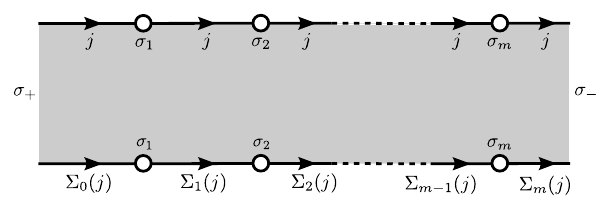}
\caption{A strip on a surface with an upper edge labelled $j$.} \label{fig:hurwitz-strip}
\end{figure}

We can encode the data given by a tuple $(\sigma_+, \sigma_1, \sigma_2, \ldots, \sigma_m; \sigma_-^{(1)}, \ldots \sigma_-^{(n-1)})$ in terms of the variables $\Sigma_0:=\sigma_+$, $\Sigma_k = \sigma_+\sigma_1\ldots \sigma_k$. This alternative tuple
\bea
(\Sigma_0, \Sigma_1, \Sigma_2, \ldots, \Sigma_m; \sigma_-^{(1)}, \ldots \sigma_-^{(n-1)}) 
\eea
directly encodes how the strips of a surface are glued, as is shown in Figure \ref{fig:hurwitz-strip}. 
Each strip with upper edges all labelled $j$ is glued to strips with lower edges labelled $\Sigma_k(j)$.
Given a tuple with permutations $\Sigma_k$, we can recover the original tuple with $\sigma_+ =\Sigma_0$, $\sigma_k = \Sigma^{-1}_{k-1}\Sigma_k$.

\begin{figure}[t]
\centering
\includegraphics[height=0.45\textheight]{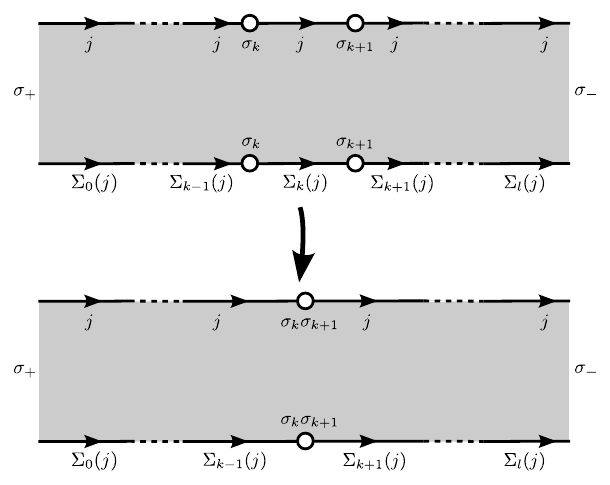}
\caption{Taking the limit of a time inequality.} \label{fig:tshrink}
\end{figure}

We consider the boundaries of the cell corresponding to the time-ordering constraints. These boundaries arise
in the limit when two time coordinates merge together.  Focusing on the (cell) codimension one boundaries of a cell,
we first take a point in the bulk of the cell at which all $l$ internal vertices have distinct time coordinates, and label the associated single-cycle permutations $\sigma_1,\ldots, \sigma_l$. This is an expanded split tuple of the graph.
For a pair of zeroes $\sigma_k$, $\sigma_{k+1}$ which cannot be commuted past each other by slide-equivalence, there is
an associated time-inequality $t_k<t_{k+1}$. As can be seen in Figure \ref{fig:tshrink}, taking the limit as $t_{k+1}\to t_k$ 
corresponds to removing the internal edges directly following the $\sigma_k$. The effect on the permutation is to replace
the consecutive cycles $\sigma_k, \sigma_{k+1}$ with the single permutation $\sigma_k\sigma_{k+1}$:
\bea 
( \sigma_+ , \ldots , \sigma_{ k-1} , \sigma_{k} , \sigma_{k+1} , \ldots ;\sigma_-^{(1)},\ldots ) 
\rightarrow ( \sigma_+ , \cdots , \sigma_{k-1} , \sigma_{k}  \sigma_{k+1} , \ldots ; \sigma_-^{(1)}, \ldots ) 
\eea
In terms of the $\Sigma_i$ description, this type of cell incidence corresponds to dropping $\Sigma_k$ from the $\Sigma$-tuple:
\bea
(\Sigma_0, \ldots, \Sigma_{k-1}, \Sigma_k, \Sigma_{k+1}, \ldots;\ \sigma_-^{(1)}, \ldots) \to (\Sigma_0, \ldots, \Sigma_{k-1}, \Sigma_{k+1}, \ldots;\ \sigma_-^{(1)}, \ldots) 
\eea

Such a contraction of the strip can in general change the genus of the strip-decomposed surface. In this case, the boundary of the cell is not a cell in $\cM_{g,n}$.
We can formulate a condition to ascertain if  a contraction of permutations gives a surface of the same genus by using the Riemann-Hurwitz formula,
\bea
2g-2+n = \sum_{k=1}^m(d-C_{\sigma_k}). \label{eq:rh2}
\eea
%where $C_{\sigma_i}$ is the number of disjoint cycles of the permutation $\sigma_i$, including trivial cycles.
The expression $(d-C_{\sigma_i})$ is the {\bf branching number} of the permutation $\sigma_i$.
If each $\sigma_i$ is a single non-trivial cycle, then a contraction of the subsequent non-disjoint cycles
$\sigma_k$ and $\sigma_{k+1}$ will preserve the genus if the sum over the branching numbers is preserved:
\bea 
( d - C_{ \sigma_{k}} ) + ( d - C_{ \sigma_{ k+1} } ) = ( d - C_{ \sigma_k \sigma_{k+1} }). \label{eq:branchings}
\eea
When this condition is satisfied, the permutation tuple resulting from the contraction defines a 
new slide equivalence class with one fewer time parameter, and so specifies a cell in $\cM_{g,n}$ with one fewer dimension.
For a given time-inequality $t_k<t_{k+1}$, there can be many different boundaries of the cell, which correspond to the different 
initial choices of the time-coordinates of the remaining interaction vertices.

As an example of when the genus-preserving condition is not satisfied, the contraction of a pair of permutations satisfying $\sigma_k\sigma_{k+1}=1$ 
cannot preserve the genus, as the right-hand-side of \eref{eq:branchings} is zero in this case and the left-hand-side is always positive. 
Another example of a genus-reducing contraction is when
$\sigma_{k} = ( 1 23 )$, $ \sigma_{k+1} = ( 234)$, and $ \sigma_{k} \sigma_{k+1} = (13) (24)$. 
Here, the branching numbers before contraction add up to four, but the product permutation has a branching number of two.
We present some examples of interaction time contractions where genus is conserved in the following subsections.

Next, we consider the boundaries of the cell corresponding to the strip-width constraint $b_j>0$. A strip can be collapsed to zero width if the label on the upper bound of the strip $j$ is not fixed by $\sigma_-$. 
From inspection of the labels on the upper and lower edges of a strip, such as in Figure \ref{fig:hurwitz-strip}, we can see that the collapse of a strip $b_j\to 0$ corresponds to replacing $j$ in each permutation $\Sigma_k$ with $\Sigma_k(j)$ when $\Sigma_k(j)\neq j$, or dropping the 1-cycle $(j)$ from $\Sigma_k$ when $\Sigma_k(j)=j$.
In other words, we remove each occurrence of $j$ from each permutation in the $\Sigma$-tuple 
\bea
(\Sigma_0, \Sigma_1, \Sigma_2, \ldots, \Sigma_m; \sigma_-^{(1)}, \ldots \sigma_-^{(n-1)}).
\eea
To find a codimension one boundary of a cell from strip-collapse,
we first take a point in the cell with distinct time coordinates for the $l$ zeroes.
We construct the associated $\Sigma$-tuple, remove each occurrence of $j$ from the tuple, relabel the tuple with labels in $\{1, \ldots, d-1\}$, and rewrite the tuple back in terms of $\sigma_k$, dropping any identity permutations appearing within the $\sigma_k$, $k=1,\ldots, l$.

As with the contraction of internal edges, the surface constructed from contracting a strip
might not have the same genus as the original surface. In the case that strip contraction preserves
the genus, then the slide-equivalence class of the new tuple specifies a new cell on the boundary of the original cell.
The Riemann-Hurwitz formula \eref{eq:rh2} relates the sum of the branching numbers of the zeroes of a tuple to the genus and number of poles of the surface. We can deduce that a strip contraction of the tuples 
$(\sigma_+, \sigma_1,\ldots,\sigma_-) \mapsto (\tilde\sigma_+, \tilde\sigma_1,\ldots,\tilde\sigma_-)$ preserves the genus of the surface
if and only if the sum over branching numbers is conserved in the contraction:
\bea
\sum_{i=1}^m (d-C_{\sigma_i}) = \sum_{i=1}^m (d-1-{C}_{\tilde{\sigma}_i}).
\eea

\subsection{Examples: \texorpdfstring{$\cM_{0,4}$}{M0,4} and \texorpdfstring{$\cM_{0,1[3]}$}{M0,13} }

We can explicitly derive the cell complex of $\cM_{g,n}$ using this procedure for some examples 
of low genus and few punctures. 
One of the simplest non-trivial examples of a moduli space is $\cM_{0,4}$, 
the space of inequivalent Riemann surfaces with four distinguishable labelled points.
It is not hard to construct this space explicitly without reference to the Nakamura cell
decomposition. Consider a base-space Riemann sphere with three labelled points, which we choose to be  $(1, e^{2i\pi/3}, e^{4i\pi/3})$. The group of biholomorphic maps on the Riemann sphere are the M\"{o}bius maps, and given a Riemann sphere with four marked points $q_1, q_2, q_3, q_4$, then there exists a unique M\"{o}bius map which takes $q_1 \mapsto 1$, $q_2 \mapsto e^{2i\pi/3}$, $q_3 \mapsto e^{4i\pi/3}$. This biholomorphism maps $q_4$ to some point $z$, with $z^3\neq 1$.
The only biholomorphism fixing three points on the Riemann sphere is the identity, 
and so this map is unique; any two Riemann spheres with four labelled points are related by a biholomorphism if and only there are M\"obius maps taking $q_4$ to the same point $z$
with $q_1 \mapsto 1$, $q_2 \mapsto e^{2i\pi/3}$, $q_3 \mapsto e^{4i\pi/3}$.
This $z$ parametrises the equivalence classes of spheres with four labelled points, and so
we deduce that the moduli space is the Riemann sphere with three punctures,
\bea
\cM_{0,4} = \mathbb{C_\infty}\backslash\{1, e^{2i\pi/3}, e^{4i\pi/3}\}.
\eea
In this subsection, we show that the cell decomposition arising from Nakamura graphs reproduces this moduli space.

\begin{figure}[t]
        \centering
\subfloat[]{\includegraphics[height=0.14\textheight]{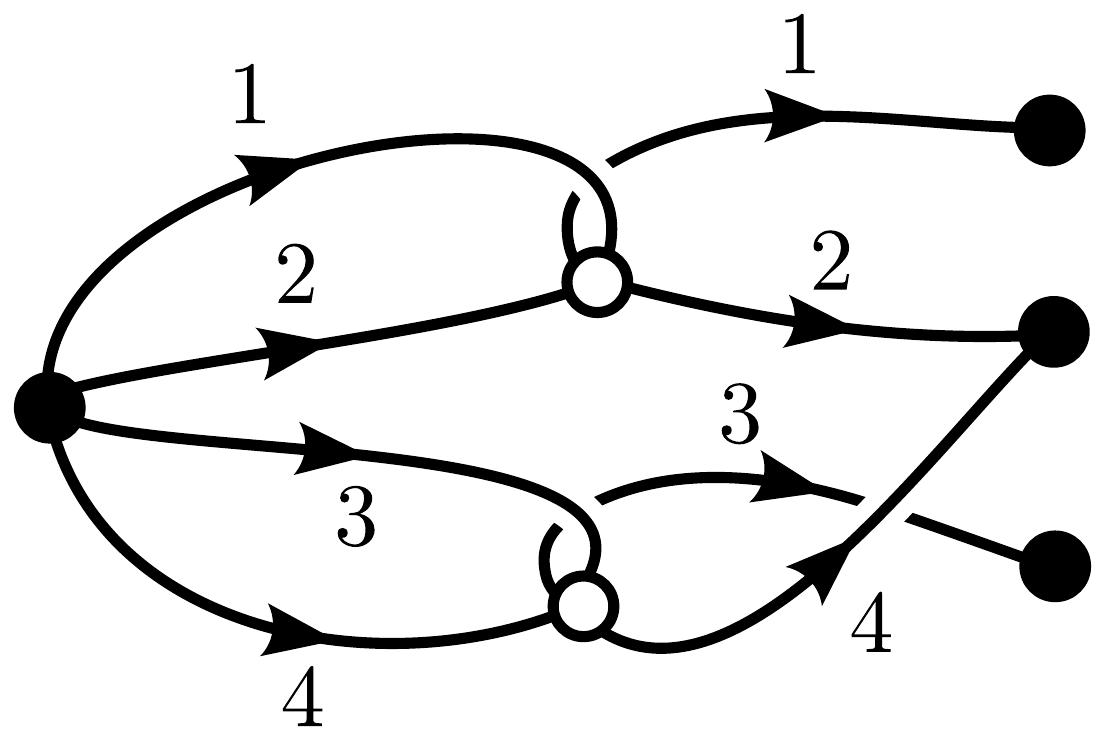}  \label{fig:g04_a}}\quad
\subfloat[]{\includegraphics[height=0.14\textheight]{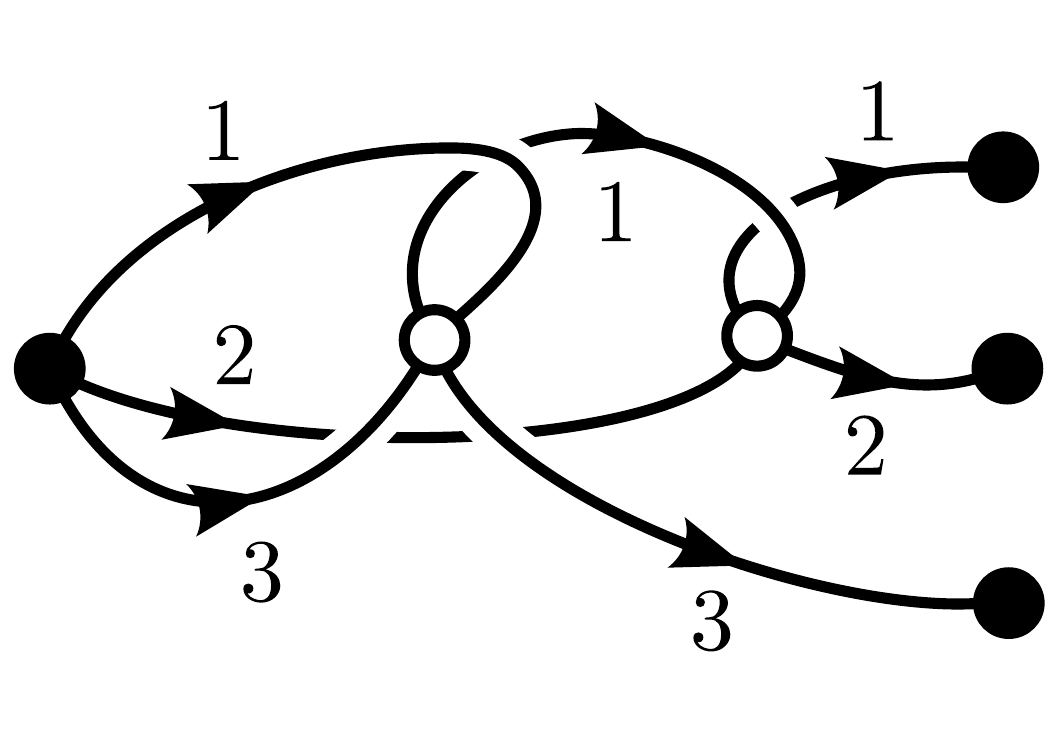}  \label{fig:g04_b}}\quad
\subfloat[]{\includegraphics[height=0.14\textheight]{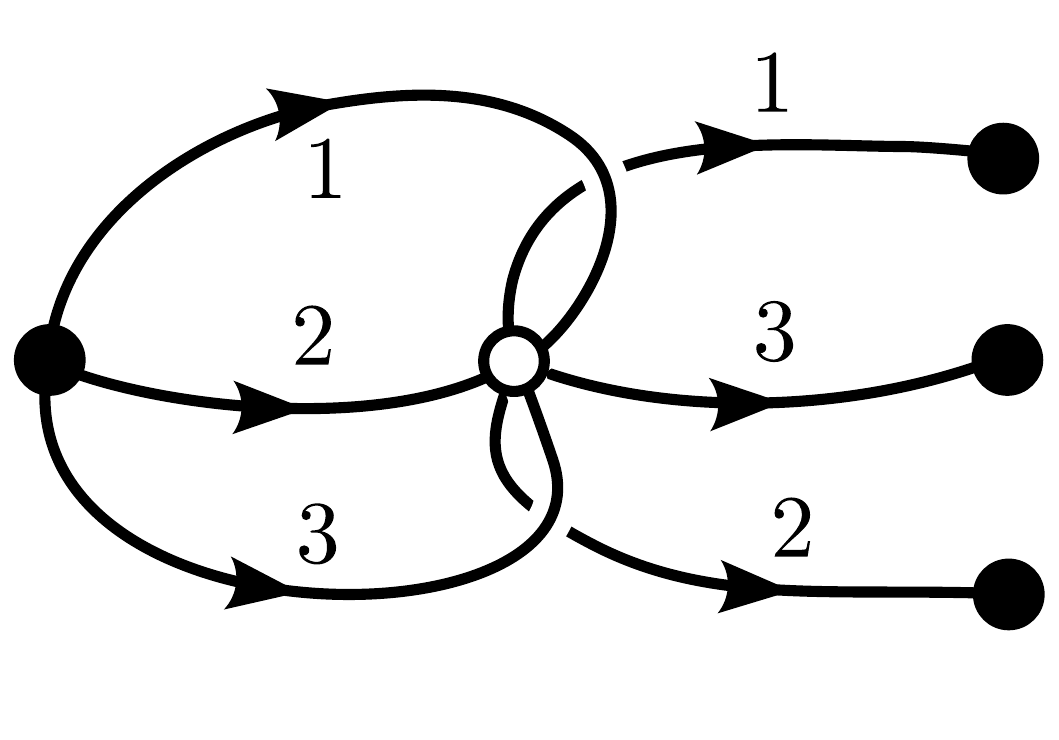}  \label{fig:g04_c}}
        \caption{The Nakamura graphs of $\cM_{0,4}$ (without pole labellings).}\label{fig:04graphs}
\end{figure}

First, we choose an arbitrary set of residues $r_1, r_2, r_3$ to assign to the outgoing poles, where $r_i<0$. There are three distinct Nakamura graphs of genus zero with four external points: these are shown in Figure \ref{fig:04graphs}.
There are several distinct possible labellings of the outgoing poles for each graph.
Each pole-labelled graph corresponds to a distinct cell in the cell decomposition of $\cM_{0,4}$.
The cells in moduli space corresponding to the same labelled graph are similar in structure, and differ only in their labellings
and in the assignments of the $r_i$. In this example, all the boundaries of the cells in $\cM_{0,4}$ will correspond to other cells in $\cM_{0,4}$, as there are no genus-reducing cell boundaries.

The graph in Figure \ref{fig:g04_a} was discussed in Section \ref{sec:splittuples}. There are three distinct pole-labelled Nakamura graphs corresponding to this unlabelled graph, and they are described by the split reduced tuples
\bea
A_1: (\sigma_+, \tau; \ \sigma_-^{(1)}, \sigma_-^{(2)}, \sigma_-^{(3)}) &=& ( (1234), (12)(34); \ (1), (24), (3)  ),\\
A_2: (\sigma_+, \tau; \ \sigma_-^{(2)}, \sigma_-^{(1)}, \sigma_-^{(3)}) &=&( (1234), (12)(34); \ (24), (1), (3)  ),\\
A_3: (\sigma_+, \tau; \ \sigma_-^{(1)}, \sigma_-^{(3)}, \sigma_-^{(2)}) &=&( (1234), (12)(34); \ (1), (3), (24)  ).
\eea
The pole-labelled graphs associated to these tuples were drawn in Figure \ref{fig:hurwitz-poles}.
The cell $A_1$ associated to the first of these tuples is a subset of $\mathbb{R}^6$, with the coordinates
\linebreak$\{ b_1, b_2, b_3, b_4; t_1, t_2\}$, subject to the constraints:
\bea
b_1 &=& |r_1|, \\
b_2+b_4 &=& |r_2|, \\
b_3 &=& |r_3|, \\
t_1 + t_2 &=& 0, \\
b_2, b_4 &>& 0.
\eea
These equations define a two-dimensional subspace of $\mathbb{R}^6$, which we can parametrise by the two variables
$b_2$ and $t_2$, subject to the relation
\bea
0< b_2 < |r_2|.
\eea
The time coordinate $t_2$ is unconstrained and can take any real value.
This means that the cell associated to this tuple is an infinite strip of width $|r_2|$.
The upper and lower boundaries of the strip correspond to the limiting values of $b_2=|r_2|$ (i.e. $b_4=0$) and $b_2=0$ respectively.
The other two cells $A_2$ and $A_3$ are defined by similar sets of equations, but with the $r_i$ interchanged, and can be parametrised by strips of width $|r_1|$ and $|r_3|$ respectively.
The pole-fixing automorphism group of the graph is trivial, and so there is no orbifolding
of the cells.

%
%\begin{figure}[t]
%\centering
%\includegraphics[width=0.4\textwidth]{c04_a}
%\caption{The 2-dimensional cell is an infinite strip, bounded by four half-lines 
%
%. and 1-dimension cells are strips and half-lines respectively. (The zero-dimension cells are points.)}\label{fig:cells}
%\end{figure}
The graph in Figure \ref{fig:g04_b} has six distinct pole labellings, given by the split tuples
\bea
B_1: (\sigma_+, \tau_1, \tau_2; \ \sigma_-^{(1)}, \sigma_-^{(2)}, \sigma_-^{(3)}) &=& ( (123), (13), (12) ; \ (1), (2), (3)  ), \label{eq:cellb1} \qquad \\
B_2: (\sigma_+, \tau_1, \tau_2; \ \sigma_-^{(2)}, \sigma_-^{(3)}, \sigma_-^{(1)}) &=& ( (123), (13), (12) ; \ (2), (3), (1)  ),\\
B_3: (\sigma_+, \tau_1, \tau_2; \ \sigma_-^{(3)}, \sigma_-^{(1)}, \sigma_-^{(2)}) &=& ( (123), (13), (12) ; \ (3), (1), (2)  ),\\
B_4: (\sigma_+, \tau_1, \tau_2; \ \sigma_-^{(1)}, \sigma_-^{(3)}, \sigma_-^{(2)}) &=& ( (123), (13), (12) ; \ (1), (3), (2)  ),\\
B_5: (\sigma_+, \tau_1, \tau_2; \ \sigma_-^{(2)}, \sigma_-^{(1)}, \sigma_-^{(3)}) &=& ( (123), (13), (12) ; \ (2), (1), (3)  ),\\
B_6: (\sigma_+, \tau_1, \tau_2; \ \sigma_-^{(3)}, \sigma_-^{(2)}, \sigma_-^{(1)}) &=& ( (123), (13), (12) ; \ (3), (2), (1)  ).
\eea
The cell $B_1$ associated to the first of these tuples is a subspace of $\mathbb{R}^5$ with coordinates $\{ b_1, b_2, b_3; t_1, t_2\}$,
defined by the constraints
\bea
b_1 &=& |r_1|, \\
b_2 &=& |r_2|, \\
b_3 &=& |r_3|, \\
t_1 + t_2 &=& 0, \\
t_1 &<& t_2.
\eea
There is only one free parameter in this cell, which we can take to be $t_2$, which satisfies
\bea
0 < t_2 < \infty.
\eea
The cell $B_1$ is a half-line, with a boundary at the point $t_2=0$. The automorphism group of the cell is trivial. The other cells $B_2,\ldots, B_6$ have similar descriptions.

Finally, the graph in Figure \ref{fig:g04_c} has two associated pole-labellings, corresponding
to the split tuples
\bea
C_1: (\sigma_+, \tau; \ \sigma_-^{(1)}, \sigma_-^{(2)}, \sigma_-^{(3)}) &=& ( (123), (132) ; \ (1), (2), (3)  ), \label{eq:0cella} \\
C_2: (\sigma_+, \tau; \ \sigma_-^{(1)}, \sigma_-^{(3)}, \sigma_-^{(2)}) &=& ( (123), (132) ; \ (1), (3), (2)  ). \label{eq:0cellb}
\eea
For the first cell $C_1$, the variables $b_1, b_2, b_3, t$ satisfy the constraints
\bea
b_1&=& |r_1|, \\
b_2&=& |r_2|, \\
b_3&=& |r_3|, \\
t &=& 0.
\eea
All four variables are fixed, so this cell is zero-dimensional. The automorphism group is trivial, and so there is no orbifolding. Similarly, $C_2$ is a zero-dimensional cell.

Collating the above, the cell decomposition of $\cM_{0,4}$ consists of three cells $A_1,A_2,A_3$ that are two-dimensional strips in parameter space, six cells $B_1,\ldots,B_6$ that are 1-dimensional half-lines, and a pair of 0-dimensional point cells $C_1,C_2$.
The incidences of these cells can be found by looking at the tuples and applying
the strip-collapsing or interaction point-collapsing algorithms. It can be shown that each of
the two boundaries of a cell $A_i$ contains a pair of half-line cells $B_j$, and that these half-lines share a boundary cell $C_k$.
We will demonstrate the procedure for one of the boundaries of the strip cell $A_1$ and a boundary
of the half-line cell $B_1$.

The cell $A_1$ is parametrised by the half-strip in $(b_2, t_2)$-space, where $0<b_2<|r_2|$,
and $t_2$ can take any value.
Consider the case when the strip width $b_2$ approaches $|r_2|$, with $t_2<0$.
In this range of time coordinates, the zero corresponding to $(34)$ appears at an earlier
time than the zero corresponding to $(12)$, and so the split tuple of points in the cell
in this range is
\bea
(\sigma_+, \sigma_1, \sigma_2; \ \sigma_-^{(1)}, \sigma_-^{(2)}, \sigma_-^{(3)}) &=& ( (1234), (34), (12); \ (1), (24), (3) ).
\eea
The limit $b_2\to|r_2|$ corresponds to taking the strip width $b_4\to 0$.
Employing the procedure for strip collapse from the previous section, we convert the tuple into
the $\Sigma$-notation:
\bea
(\Sigma_0, \Sigma_1, \Sigma_2; \ \sigma_-^{(1)}, \sigma_-^{(2)}, \sigma_-^{(3)}) =
((1234), (124)(3), (1)(24)(3); (1),(24),(3)).\qquad
\eea
Here, we have written out the 1-cycles explicitly. Collapsing the strip with upper edges labelled 4 corresponds to removing the integer 4 from each cycle in the $\Sigma$-tuple:
\bea
(\Sigma_0, \Sigma_1, \Sigma_2; \ \sigma_-^{(1)}, \sigma_-^{(2)}, \sigma_-^{(3)}) =
((123), (12)(3), (1)(2)(3); (1), (2), (3)).\qquad
\eea
Converting this back into the $\sigma$-notation, we find the split tuple
\bea
(\sigma_+, \sigma_1, \sigma_2; \ \sigma_-^{(1)}, \sigma_-^{(2)}, \sigma_-^{(3)}) 
= ((123), (13), (12); (1),(2),(3)). \label{eq:1cell}
\eea
This is the split tuple of the cell $B_1$, given in \eref{eq:cellb1}.
We deduce that the $b_2=|r_2|, t_2<0$ boundary of the cell $A_1$ is the cell $B_1$.
A similar procedure can be applied to the $b_2=|r_2|, t_2>0$ boundary, the $b_2=0, t_2<0$ boundary, 
and the $b_2=0, t_2>0$ boundary of the cell. We recover the split tuples of the half-line cells 
$B_3$, $B_5$, and $B_6$ respectively.

The cell $B_1$ is a half-line, corresponding to the split tuple \eref{eq:1cell}, 
and parametrised by the time coordinate $\tilde{t}_2$ of the interaction point associated
to the cycle $(12)$, where $\tilde{t}_2>0$. (Note that this time coordinate is not the same as the time coordinate on the cell $A_1$ given above, as the ordering of the interaction vertices has been interchanged.)
Taking the limit $\tilde{t}_2\to 0$ while $\tilde{t}_1+\tilde{t}_2=0$ is held fixed corresponds to merging the two interaction points together.
The corresponding new interaction point is described by the product of the permutations of the original two points, as was shown in the previous subsection.
The merging of the two points $\sigma_1=(13)$ and $\sigma_2=(12)$ generates the new split tuple
\bea
(\sigma_+, \tau; \ \sigma_-^{(1)}, \sigma_-^{(2)}, \sigma_-^{(3)}) =
( (123), (132); \ (1), (2), (3) )
\eea
which is the split tuple \eref{eq:0cella} of the 0-dimensional cell $C_1$.

\begin{figure}[t]
        \centering
\subfloat[]{\includegraphics[height=0.155\textheight]{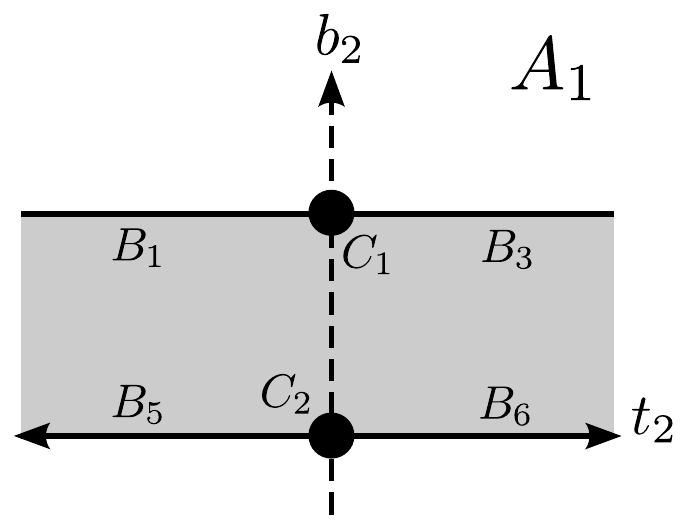}  \label{fig:cells-a04a}}\quad
\subfloat[]{\includegraphics[height=0.155\textheight]{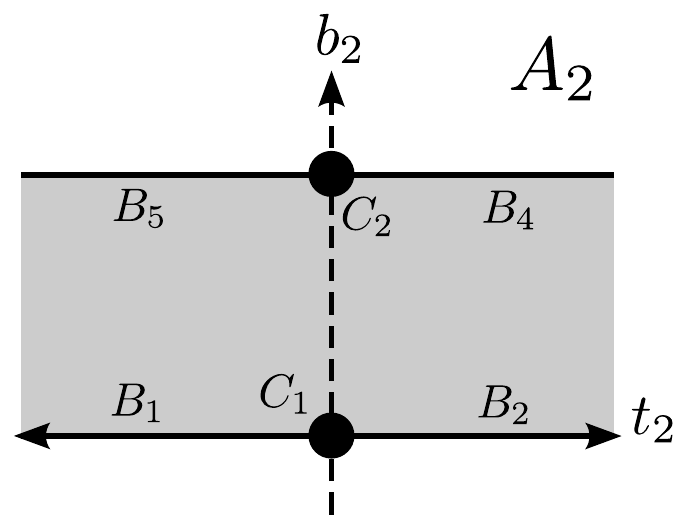}  \label{fig:cells-a04b}}\quad
\subfloat[]{\includegraphics[height=0.155\textheight]{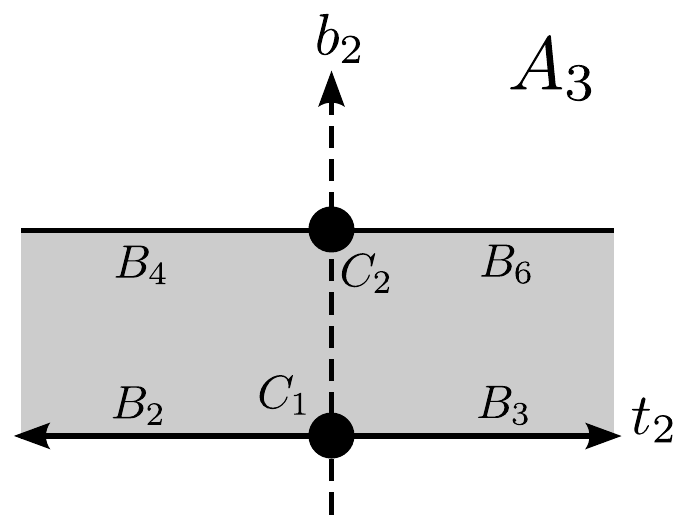}  \label{fig:cells-a04c}}
        \caption{The 2-dimensional cells $A_1$, $A_2$, $A_3$, and their incidences with lower-dimensional cells.}\label{fig:cells-a}
\end{figure}

\begin{figure}[t]
\centering
\includegraphics[width=0.35\textwidth]{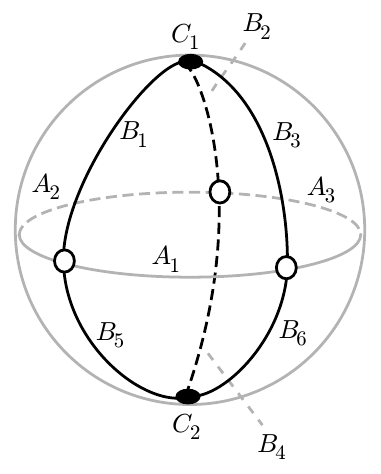}
\caption{The cell decomposition of $\cM_{0,4}$.} \label{fig:cells-m04}
\end{figure}

By performing a similar analysis on each cell $A_1$, $A_2$, $A_3$ in turn, we find that each cell borders four distinct 1-cells and two distinct 0-cells. A diagram showing the cells and their incidences is given in Figure \ref{fig:cells-a}.
These cells glue together to form a sphere with three punctures, as shown in Figure \ref{fig:cells-m04}, where the positive and negative infinities of the strips have been homeomorphically mapped to the white vertices of the diagram.
This picture agrees with the description of $\cM_{0,4}$ as a three-punctured sphere given at the beginning of this section.

\begin{figure}[t]
\centering
\includegraphics[height=0.2\textheight]{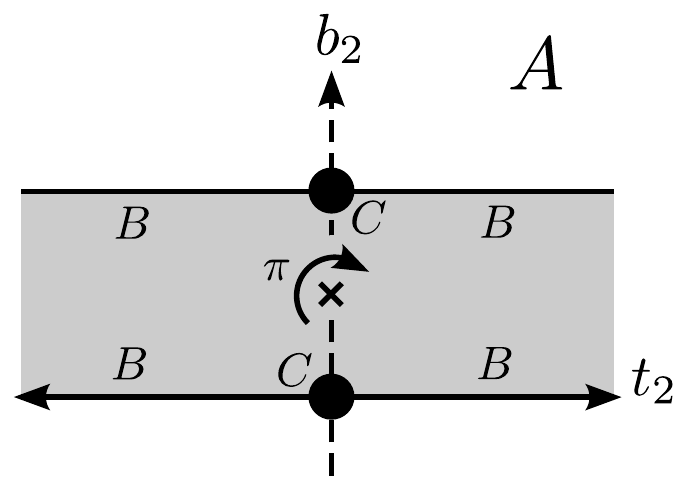}
\caption{The 2-cell $A$ in the cell decomposition of $\cM_{0,1[3]}$.} \label{fig:cells-a013}
\end{figure}

We can also obtain the cell decomposition of $\cM_{0,1[3]}$ by using this cell decomposition of $\cM_{0,4}$.
First, set $r_1=r_2=r_3=r$ for some negative real $r$.
The split tuples associated to the cells $C_1, C_2$ in $\cM_{0,4}$ differ only by a rearrangement of the cycles representing the poles, and so these cells correspond to the same cell $C$ in $\cM_{0,1[3]}$, described by the tuple
\bea
(\sigma_+, \tau, \sigma_-) = ((123), (132), ()).
\eea
Similarly, there is a single 1-cell $B$ in $\cM_{0,1[3]}$ associated to the six 1-cells $B_1,\ldots, B_6$ of $\cM_{0,4}$.
There is also a single cell $A$ in $\cM_{0,1[3]}$, corresponding to the three cells $A_1,A_2,A_3$,
with the reduced tuple
\bea
(\sigma_+, \tau, \sigma_-) = ((1234), (12)(34), (24)).
\eea
The covering space $\cB(\cG)$ of the cell $A$ is defined by the constraints
\bea
b_1 &=& |r|, \\
b_2+b_4 &=& |r|, \\
b_3 &=& |r|, \\
t_1 + t_2 &=& 0, \\
b_2, b_4 &>& 0.
\eea
The tuple has the non-trivial automorphism group $\mathbb{Z}_2$ generated by the permutation $(13)(24)$. This acts on the parameter space by interchanging the time coordinates of the zeroes and strip widths $t_1\leftrightarrow t_2$, $b_1\leftrightarrow b_3$, $b_2\leftrightarrow b_4$.
This acts on the parametrising strip  $-\infty<t_2<\infty$, $0<b_2<|r|$ as a rotation by $\pi$ 
about the point $ t_2=0, b_2=|r|/2$.
The cell $A$ in $\cM_{0,1[3]}$ is the quotient space $\cB(\cG)/\mathbb{Z}_2$.
This is shown in Figure \ref{fig:cells-a013} 

The moduli space $\cM_{0,1[3]}$ can be visualised as a quotienting of $\cM_{0,4}$.
In the M\"obius maps description, a pair of points on the three-punctured sphere represent
the same point in $\cM_{0,1[3]}$ if they are related by a M\"obius map which preserves the set
of three punctures $\{1, e^{2\pi i/3}, e^{4\pi i/3}\}$. The set of M\"obius maps which preserve
these three points is the group $S_3$, and acts by rotations on the Riemann sphere shown in Figure \ref{fig:cells-m04}. We deduce that
\bea
\cM_{0,1[3]} = (\mathbb{C_\infty}\backslash\{1, e^{2i\pi/3}, e^{4i\pi/3}\})/ S_3.
\eea
In terms of the cells decomposition of $\cM_{0,4}$ shown in Figure \ref{fig:cells-a}, the quotienting by $S_3$ identifies the 2-cells labelled $A_1,A_2,A_3$,
the 1-cells labelled $B_1,\ldots, B_6$, and the 0-cells labelled $C_1, C_2$.
The 2-cell of $\cM_{0,1[3]}$ acquires a non-trivial automorphism group $\mathbb{Z}_2$ in this quotienting.

%$
\subsection{Example: \texorpdfstring{$\cM_{1,2}$}{M1,2} }

\begin{figure}[t]
        \centering
\subfloat[]{\includegraphics[width=0.35\textwidth]{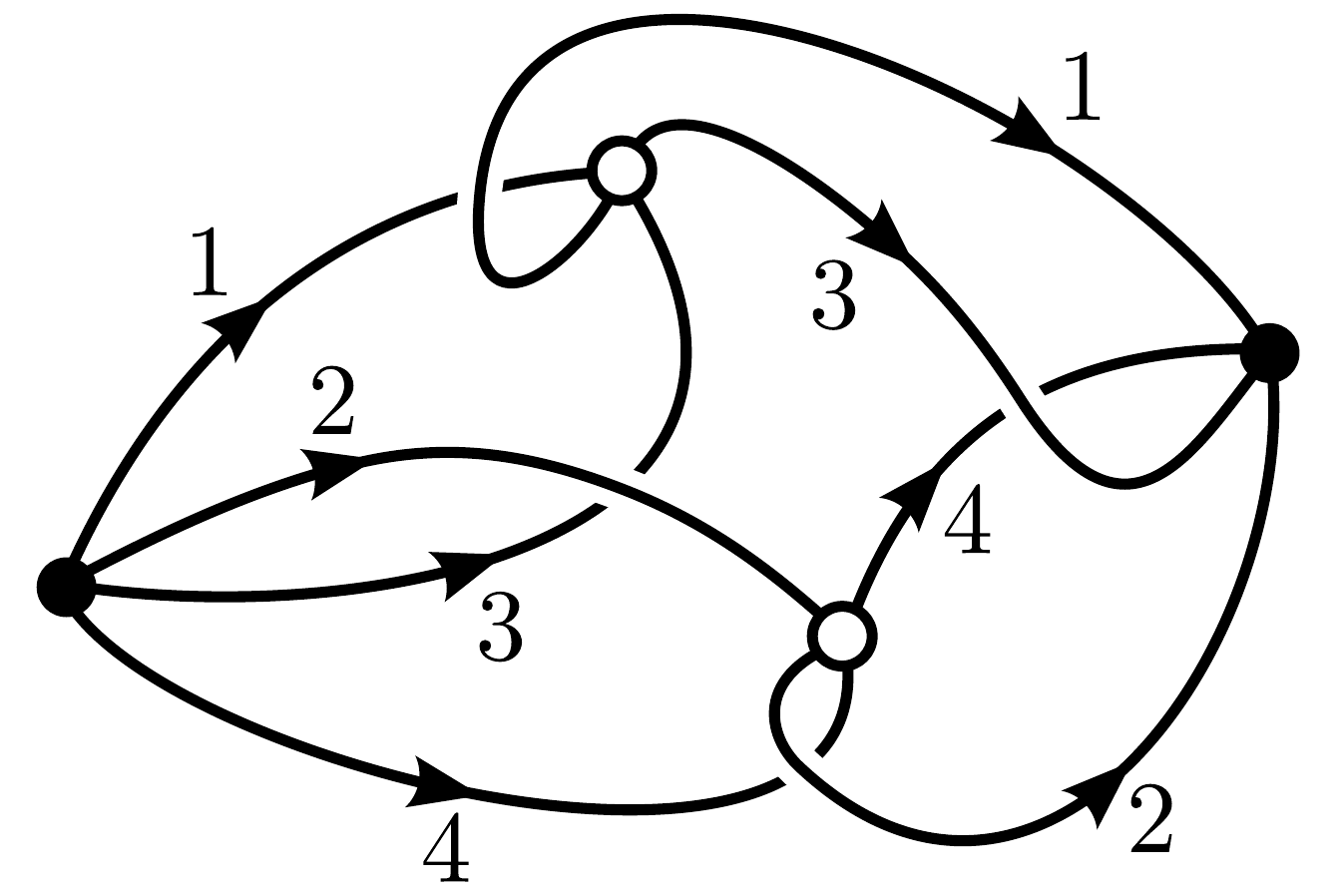} \label{fig:cells-graph12a}}\qquad
\subfloat[]{\includegraphics[width=0.35\textwidth]{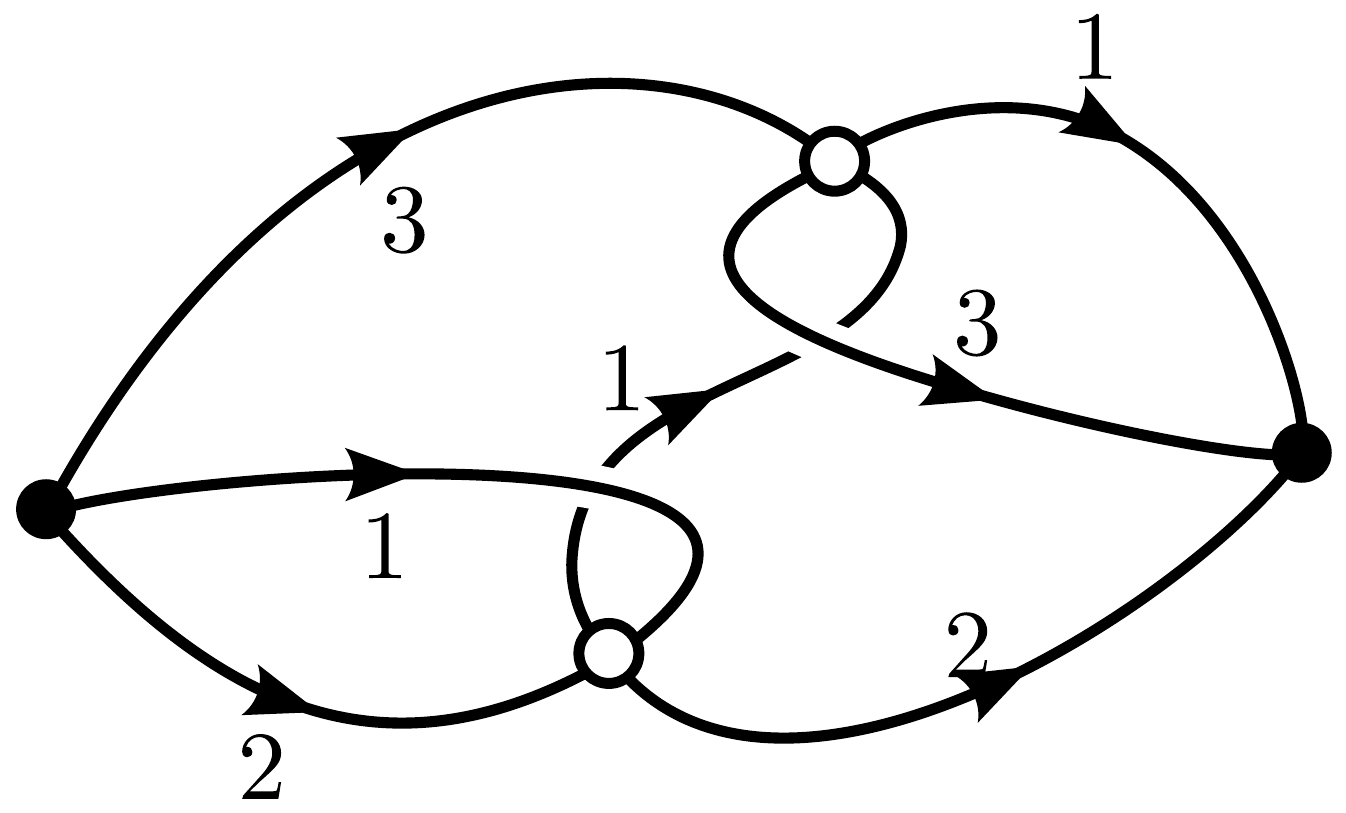} \label{fig:cells-graph12b}} \qquad
\subfloat[]{\includegraphics[width=0.35\textwidth]{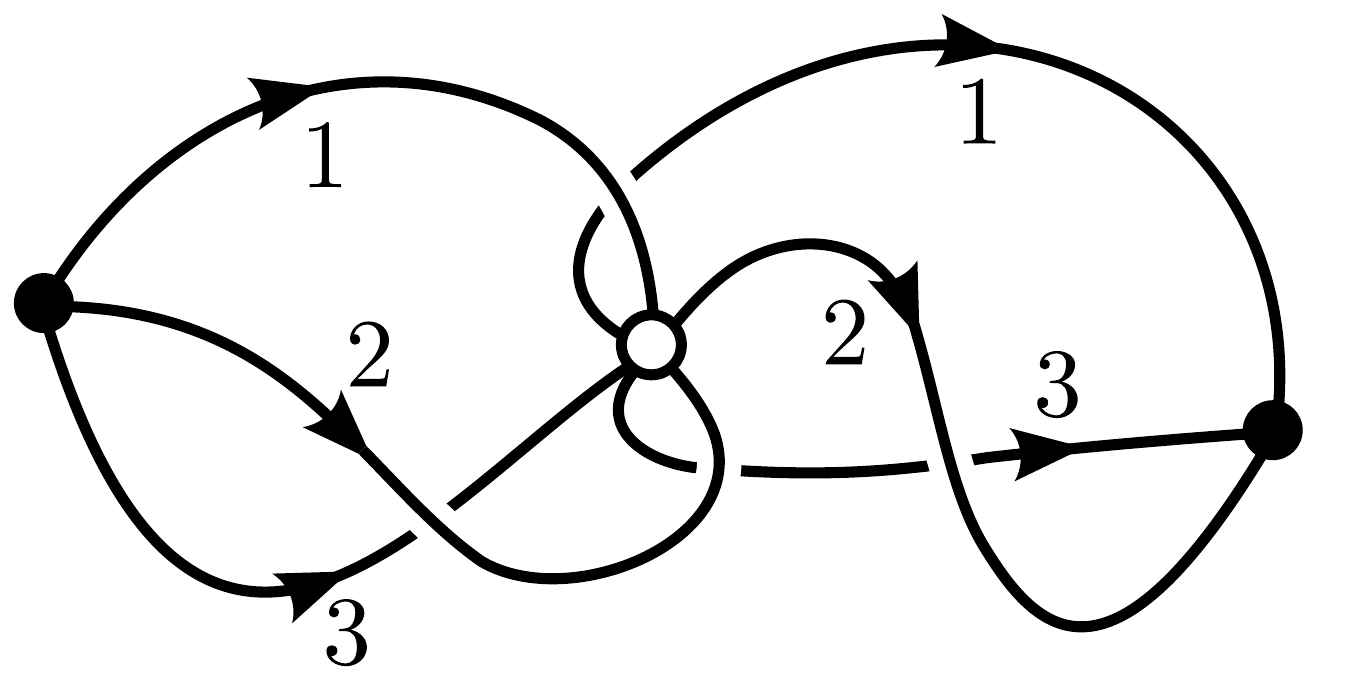} \label{fig:cells-graph12c}}\qquad
\subfloat[]{\includegraphics[width=0.35\textwidth]{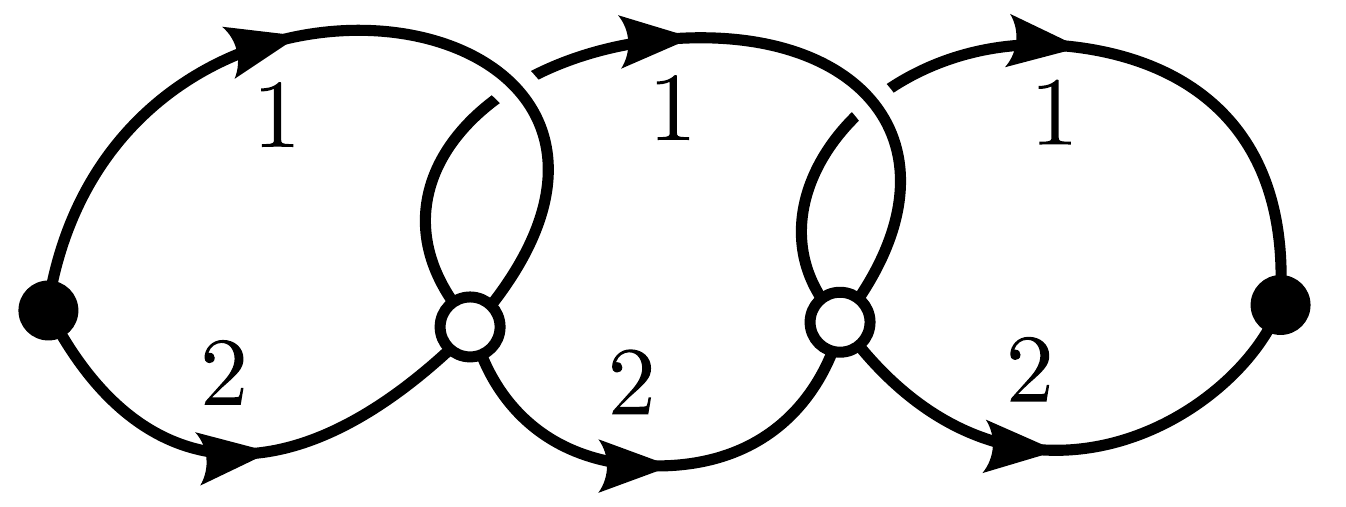} \label{fig:cells-graph12d}}
        \caption{The Nakamura graphs of $\cM_{1,2}^{\mathrm{fix}}$.}\label{fig:cells-graph12}
\end{figure}

In this subsection we present another example of an explicit low-dimensional cell decomposition
of a moduli space.
There are four Nakamura graphs with $g=1$ and $n=2$, which are shown in Figure \ref{fig:cells-graph12}.
These graphs correspond to the cells of the four-dimensional moduli space $\cM_{1,2}$.
As there is one incoming and one outgoing pole for each graph, there is just one possible labelling of the outgoing pole for each graph, and so each graph corresponds to exactly one cell in the cell decomposition of moduli space. We refer to these cells as cell $A$, $B$, $C$, and $D$ respectively.
Unlike the previous example of $\cM_{0,4}$, the automorphism groups of these graphs are non-trivial, so $\cM_{1,2}$ is not a manifold. Also, as this is a moduli space of Riemann surfaces with non-zero genus, there can be cell boundaries that correspond to surfaces of reduced genus, and so are not contained within the moduli space $\cM_{1,2}$.
The pole-permuting and pole-fixing automorphism groups of the graphs coincide, and so $\cM_{1,[1]1}$ and $\cM_{1,2}$ are identical.

The four cells of  $\cM_{1,2}$ are described by the tuples
\bea
A:(\sigma_+, \tau;\ \sigma_-) &=& ((1234), (13)(24);\ (1234)), \label{eq:cellA}\\ 
B:(\sigma_+, \tau_1, \tau_2; \sigma_-) &=& ((123), (12), (13); \ (123)),\\
C:(\sigma_+, \tau; \ \sigma_-)  &=& ((123), (123); \ (123)),\\
D:(\sigma_+, \tau_1, \tau_2; \ \sigma_-) &=& ((12), (12), (12);\ (12)).
\eea
Cell $A$ is constructed by quotienting a subspace of $\{(b_1, b_2, b_3, b_4; t_1, t_2)\}$
with $\mathbb{Z}_4$, the automorphism group of the graph. The strip widths and time coordinates of points in $A$ satisfy the relations
\bea
b_1+b_2+b_3+b_4 &=& |r|, \ret 
b_i&>&0, \qquad i=1,2,3,4 \ret 
t_1 + t_2 &=& 0.
\eea
The cell is four-dimensional, as $b_4$ and $t_1$ can be written in terms of the variables $(b_1, b_2, b_3; t_2)$ which satisfy
\bea
b_i&>&0, \qquad i=1,2,3,4 \nonumber \\
b_1+b_2+b_3 &<& |r| \label{eq:3-simplex}
\eea
with $t_2$ taking any real value.
The generator $a$ of the automorphism group $\Afix=\mathbb{Z}_4$ acts on the edge labels as the permutation $(1234)$ and acts on the time coordinates as the permutation $(12)$.
Geometrically, this cell is the direct product of a 3-simplex (tetrahedron) and the real line, with a quotienting $\mathbb{Z}_4$ action on the tetrahedron and a $\mathbb{Z}_2$ action on the real line.

The automorphism group of cell $A$ identifies the four boundaries of the parameter space at $b_j=0$, so cell $A$ has a single codimension one boundary.
To determine the cell on this boundary, we first split up the time coordinates of the two vertices to generate the tuple 
\bea
(\sigma_+, \sigma_1, \sigma_2; \ \sigma_-) = ((1234), (13), (24);\ (1234)),
\eea
and employ the procedure described in subsection \ref{sec:bdy} to take the strip width $b_4\to 0$.
The tuples generated at each step are:
\bea
\overset{\Sigma}{\longrightarrow} (\Sigma_0, \Sigma_1, \Sigma_2; \ \sigma_-) &=& ((1234), (12)(34), (1432);\ (1234)) \\
\overset{-4}{\longrightarrow} (\Sigma_0, \Sigma_1, \Sigma_2; \ \sigma_-) &=& ((123), (12)(3), (132);\ (123)) \\
\overset{\sigma}{\longrightarrow} (\sigma_+, \sigma_1, \sigma_2; \ \sigma_-) &=& ((123), (13), (23);\ (123)) \\
\overset{\mathrm{relabel}}{\longrightarrow} (\sigma_+, \sigma_1, \sigma_2; \ \sigma_-) &=& ((123), (12), (13);\ (123)).
\eea
In the last step, we have relabelled the tuple by acting on the tuple by conjugation with
$(123)$.
We see that the cell $B$ lies on the boundary of $A$.

Cell $B$ is described by the variables $(b_1, b_2, b_3; t_1, t_2)$ which satisfy
\bea
b_1+b_2+b_3 &=& |r| \\
b_1,b_2,b_3 &>& 0, \\
t_1+t_2 &=& 0 \\
t_1&<&t_2
\eea
The cell is three-dimensional and can be parametrised by the variables $(b_1, b_2; t_2)$ satisfying
\bea
b_1, b_2, t_2 &>& 0 \\
b_1+b_2 &<& |r| \\
t_2&>&0.
\eea
The automorphism group $\Afix$ of the graph is trivial. Geometrically, the cell is the direct product of a 2-simplex (triangle) and a half-line.
The cell has four boundaries of the form $b_1=0$, $b_2=0$, $b_3=0$, and $t_2=0$.

The boundary $t_2\to 0$ of cell $B$ corresponds to the merging of the two interaction vertices; it can be seen that this gives the cell $C$ of the graph with a single interior vertex.
Collapsing either the strip width $b_3\to 0$ or the strip width $b_2\to 0$ leads to the tuple
\bea
(\sigma_+, \sigma_1,\sigma_2; \ \sigma_-) = ((12), (12), (12);\ (12)),
\eea
hence two of the boundaries of $B$ are the cell $D$.
However, collapsing the strip $b_1\to 0$ leads to a different tuple: we have
\bea
(\sigma_+, \sigma_1, \sigma_2; \sigma_-) &=& ((123), (12), (13); \ (123)) \\
\overset{\Sigma}{\longrightarrow} (\Sigma_0, \Sigma_1, \Sigma_2; \ \sigma_-) &=& ((123), (1)(23), (132);\ (123)) \\
\overset{-1}{\longrightarrow} (\Sigma_0, \Sigma_1, \Sigma_2; \ \sigma_-) &=& ((23), (23), (23);\ (23)) \\
\overset{\mathrm{relabel}}{\longrightarrow} (\Sigma_0, \Sigma_1, \Sigma_2; \ \sigma_-) &=& ((12), (12), (12);\ (12)) \\
\overset{\mathrm{\sigma}}{\longrightarrow} (\sigma_+, \sigma_1, \sigma_2; \ \sigma_-) &=& ((12), (), ();\ (12)). \\
\overset{\mathrm{remove}\ ()}{\longrightarrow} (\sigma_+; \ \sigma_-) &=& ((12);\ (12)).
\eea
This last tuple does not describe a cell in $\cM_{1,2}$.
We can interpret the tuple as describing a degenerate Nakamura graph, representing the single
point in the trivial moduli space $\cM_{0,2}$. 
This means that the boundary $b_1=0$ of cell $B$ is not part of the decomposition of $\cM_{1,2}$.

Cell $C$ is a two-dimensional simplex, parametrised by $b_1$ and $b_2$ satisfying
\bea
b_1, b_2 &>& 0, \\
b_1+b_2 &<& |r|,
\eea
under the quotienting group $\mathbb{Z}_3$ which acts like a $2\pi/3$ rotation on the simplex. The boundary of this cell corresponds to the degenerate tuple $((12);(12))$, so is not a part of the moduli space.
Cell $D$ is also two-dimensional and is parametrised by $b_1$ and $t_2$ satisfying
\bea
0<&b_1&<|r| \\
t_2&>&0,
\eea
under a $\mathbb{Z}_2$ quotienting. Geometrically, this cell is a half-strip of width $|r|$ with a $\mathbb{Z}_2$ quotienting associated to a reflection in the line $b_2=|r|/2$.
The boundaries $b_1=0$ and $t_2=0$ again correspond to the degenerate tuple $((12);(12))$,
and are not part of the moduli space.

\subsection{Cells in Teichm\"uller space}\label{sec:teich}

We have described how Nakamura graphs give a cell decomposition of moduli space $ \cM_{ g , n }$ 
with labelled points. We will now explain the action of  the mapping class group and show how  it acts on the 
Nakamura graphs to  give  a cell decomposition of Teichm\"uller space.
Let $\Sigma_{g,n}$ be a topological surface of genus $g$ with $n$ labelled points.
A marked complex structure on $\Sigma_{g,n}$ is a triple $(X,P_i,\phi)$ in which
$X$ is a Riemann surface with labelled points $P_i$ and $\phi:\Sigma_{g,n}\to X$ is a homeomorphism
which maps the labelled points of $\Sigma_{g,n}$ to the respective labelled points $P_i$ of $X$.
A pair of marked complex structures $(X_1,P_i^{(1)},\phi_1)$ and $(X_2,P_i^{(2)},\phi_2)$ are Teichm\"uller-equivalent if there exists a biholomorphism $f:X_1\to X_2$ such that $f\circ\phi_1$ and $\phi_2$ are isotopic, i.e. related by a continuous path of homeomorphisms
which fix the labelled points. 
The set of Teichm\"uller-equivalence classes of marked complex structures on $\Sigma_{g,n}$ is the {\it Teichm\"uller space} $\cT_{g,n}$.
More discussion on Teichm\"uller space and its relation to moduli space and mapping class groups is given in \cite{thurston, mcg}.

It was stated in \cite{nakamura} that the `marked graphs' give a cell decomposition of Teichm\"uller space, and that the action of the mapping class group preserves these cells, giving a cell decomposition of moduli space. In this section, we confirm this claim, showing that each point in the Teichm\"uller space $\cT_{g,n}$ corresponds to a Nakamura graph $\bar\cG$, an embedding of this graph on a surface $\Sigma_{g,n}$, and some strip widths $b_j$ and interaction times $t_k$.
Each cell is specified by a pair $(\bar\cG, \hat\psi)$, where $\hat\psi$ is a graph embedding into $\Sigma_{g,n}$, and this cell is in one-to-one correspondence with the parameter space $\cB(\bar\cG)$ of the graph.

On choosing a set of residues $r_1,r_2,\ldots r_n$, any Riemann surface $X$ has a Nakamura strip decomposition $(\bar\cG, b_j,t_k)$, which is unique up to relabellings of the parameters. 
Inverting the complex structure marking $\phi:\Sigma_{g,n}\to X$ gives a strip embedding $\psi:X\to \Sigma_{g,n}$, which restricts to a graph embedding on the boundaries of the strips $\hat\psi:\cG \to \Sigma_{g,n}$. A graph embedding $\hat\psi$ and a strip decomposition is enough to reconstruct
a strip embedding $\psi$ up to isotopy.

We can rephrase the above definition of Teichm\"uller space in terms of Nakamura graph embeddings and strip decompositions.
Teichm\"uller space is the set of equivalence classes of tuples of the form $(\bar\cG, \hat\psi, b_j, t_k)$ consisting of a pole-labelled graph, a graph embedding, a set of strip widths, and a set of time coordinates. A pair of tuples  $(\bar\cG^{(1)}, \hat\psi^{(1)}, b_j^{(1)}, t_k^{(1)})$ and  $(\bar\cG^{(2)}, \hat\psi^{(2)}, b_j^{(2)}, t_k^{(2)})$  are Teichm\"uller equivalent if and only if:
\begin{itemize}
\item The graphs are identical: $\bar\cG^{(1)}=\bar\cG^{(2)}$;
\item There exists an automorphism $g\in \mathrm{Aut}(\bar\cG^{(1)})$ which acts on the parameters
of the graph as $\{ b_j^{(1)}\} \mapsto\{  b_j^{(2)} \} $, $ \{ t_k^{(1)}\} \mapsto\{  t_k^{(2)} \} $;
\item The graph embeddings $\hat\psi^{(1)}$ and $\hat\psi^{(2)}\circ g$ are isotopic.
\end{itemize}

In Section \ref{sec:cells}, we constructed the set $\cB(\bar\cG)$ parametrising the possible strip decompositions $\{(b_j, t_k)\}$ of Riemann surfaces for a given graph. 
In general, there are distinct points in $\cB(\bar\cG)$ that correspond to the same Riemann surface. The number of points in $\cB(\bar\cG)$ that correspond to the same Riemann surface
$X$ is the order of the {\bf biholomorphism group} $\mathrm{Bi}(X)$, which is the group of biholomorphisms from the surface $X$ to itself.\footnote{This group is sometimes called the {\it automorphism group} $\mathrm{Aut}(X)$  of the Riemann surface in the literature.}

\begin{figure}[t]
\centering
\includegraphics[width=0.90\textwidth]{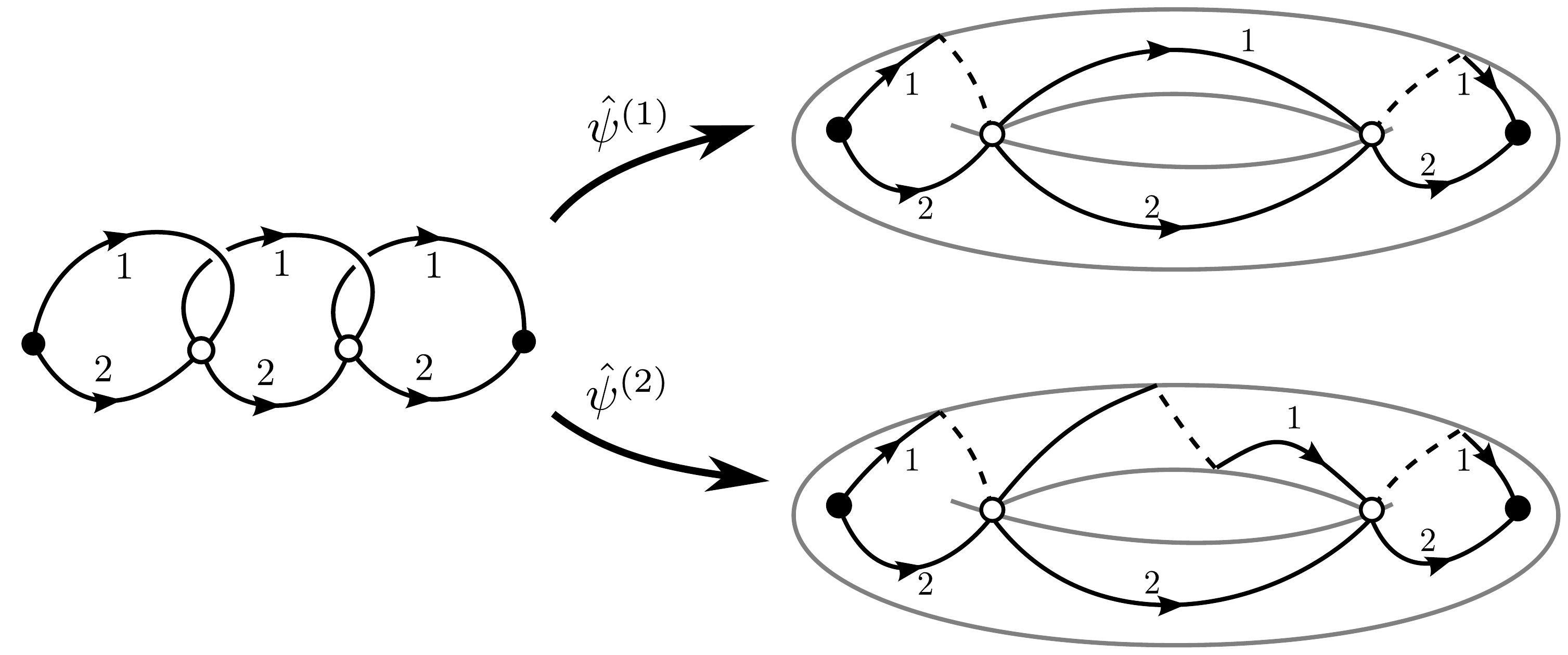}
\caption{Two embeddings of a graph which cannot be identified by isotopy.}\label{fig:teich1}
\end{figure}

For a given graph $\bar\cG$ and graph embedding $\hat\psi:\bar\cG\to\Sigma_{g,n}$,
we can show that a pair of strip decompositions of the graph $(b_j^{(1)}, t_k^{(1)})$ and $(b_j^{(2)}, t_k^{(2)})$ in $\cB(\bar\cG)$ can only correspond to Teichm\"uller-equivalent points when the parameters are identical.
For Teichm\"uller-equivalence, these strip decompositions must correspond to the same Riemann surface $X$, and there must exist a graph automorphism $g$ with the property that $\hat\psi$ and $\hat\psi\circ g$ are isotopic. 
This graph automorphism $g$ acts on the parameters of the strips as a permutation, and so extends to a biholomorphism $f_g:X\to X$. 
The graph embedding $\hat\psi$ can be extended to a strip embedding $\psi:X\to \Sigma_{g,n}$, 
and so $\psi$ and $\psi\circ f_g$ are isotopic embeddings of the strips into $\Sigma_{g,n}$, which implies that $f_g$ is isotopic to the identity.
However, it was shown by Hurwitz that non-trivial biholomorphisms of hyperbolic Riemann surfaces are not isotopic to the identity \cite{Hurwitz}. 
We deduce that $f_g$ is trivial, the graph automorphism $g$ is trivial, and so the parameters
must satisfy $b_j^{(1)}=b_j^{(2)}$, $t_k^{(1)}=t_k^{(2)}$.

This confirms that the set $\cB(\bar\cG)$, together with a graph embedding $\hat\psi$, is in one-to-one correspondence with a subset of Teichm\"uller space. 
We write $\cB(\bar\cG, \hat\psi)\equiv\cB(\bar\cG)$ for the set of points in Teichm\"uller space corresponding to a graph $\bar\cG$ and a marking $\hat\psi:\bar\cG\to \Sigma_{g,n}$.
A pair of graph markings $\hat\psi^{(1)}$ and $\hat\psi^{(2)}$ 
are equivalent if there is a graph automorphism $g\in\mathrm{Aut}(\bar\cG)$ such that
$\hat\psi^{(1)}$ and $\hat\psi^{(2)}\circ g$ are isotopic.
The set of all Nakamura graphs $\{\bar\cG\}$, and the set of all inequivalent graph embeddings $\hat\psi$ for each  graph $\bar\cG$, gives a cell decomposition of Teichm\"uller space $\cT_{g,n}$.

The mapping class group $\Gamma_{g,n}$ acts on the set of graph embeddings in a well-defined way up to isotopy, and any two markings of the same graph can be related by some element of the mapping class group. 
This shows explicitly from the Nakamura graphs picture that the quotient space of Teichm\"uller space $\cT_{g,n}$  under the mapping class group $\Gamma_{g,n}$ is the moduli space  $\cM_{g,n}$.

\begin{figure}[t]
\centering
\includegraphics[width=0.95\textwidth]{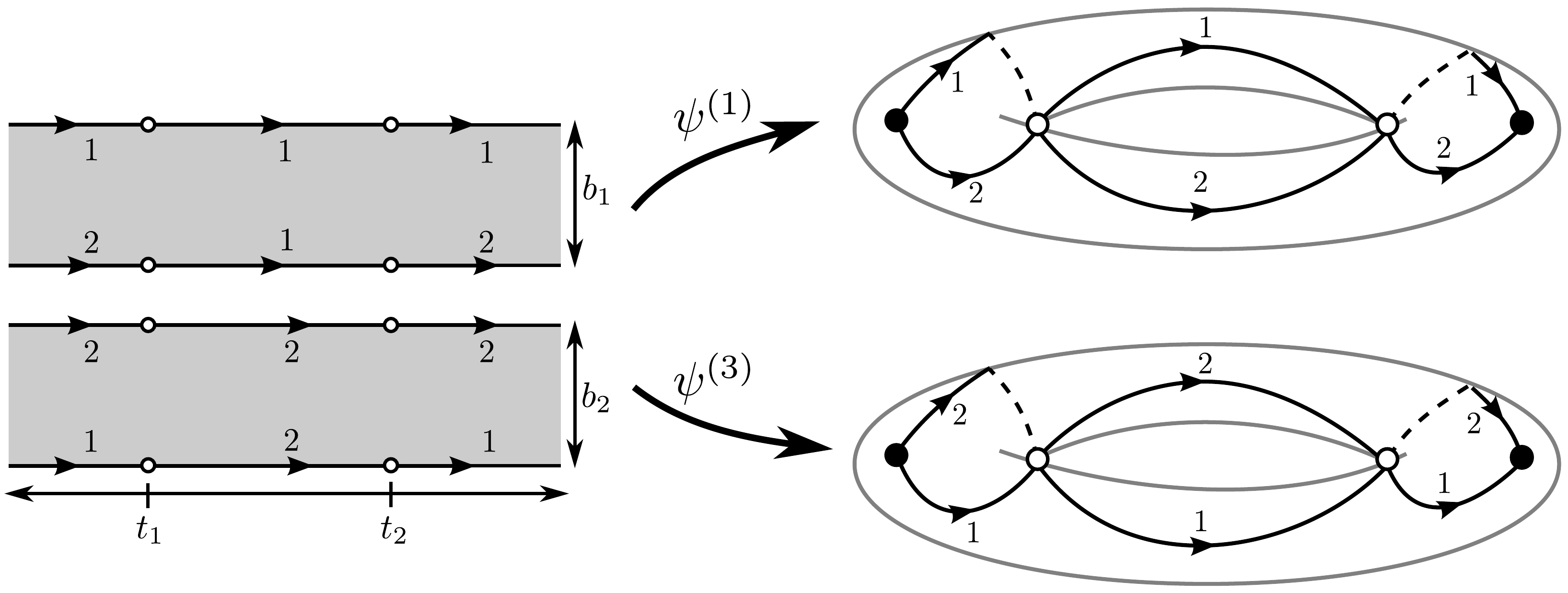}
\caption{Two embeddings of the strips into the surface, which are Teichm\"uller-equivalent if and only if the strip widths are equal.}\label{fig:teich2}
\end{figure}

We conclude this section with some examples of embeddings of a graph which was considered in the example in the previous section, determined by the tuple
\bea
(\sigma_+, \tau_1, \tau_2; \sigma_-) = ((12),(12),(12);(12)).
\eea
In Figure \ref{fig:teich1}, we have shown two embeddings $\hat\psi^{(1)}$ and $\hat\psi^{(2)}$ of the graph into the two-punctured torus that cannot be isotopic. These two markings correspond to different cells $\cB(\bar\cG, \hat\psi^{(1)})$ and $\cB(\bar\cG, \hat\psi^{(2)})$ in Teichm\"uller space. In Figure \ref{fig:teich2}, we have given two examples of strip embeddings $\psi^{(1)}$ and $\psi^{(3)}$ into the two-punctured torus. These embeddings of the strip decomposition correspond to the same point in Teichm\"uller space if and only if the strip widths are equal. 
The restriction of these strip embeddings to the graphs are $\hat\psi^{(1)}$ and $\hat\psi^{(3)}$,
which are related by a graph automorphism, and so correspond to the same Teichm\"uller space cell $\cB(\bar\cG, \hat\psi^{(1)})$. This cell is parametrised by a strip width $b_1\in(0,r)$ and an interaction time $t_2>0$, which is a semi-infinite strip.

\section{Summary}

We have provided an extension of the $S_d$ tuples description of  $\cM_{g,1[n-1]}$ to a description of  $\cM_{g,n}$.
This lead to an explicit, precise description of each cell in the Nakamura cell decomposition of $\cM_{g,n}$ and $\cM_{g,1[n-1]}$ by generating a quotiented convex polytope for each tuple.
We demonstrated this procedure explicitly for the cell decompositions of the moduli spaces $\cM_{0,4}$, $\cM_{0,1[3]}$, and $\cM_{1,2}$, and verified that these cell decompositions matched the known descriptions
of these moduli spaces.
We also gave an extension of the Nakamura cell decomposition to Teichm\"uller space, which showed explicitly that the Nakamura cell decomposition is invariant under the action of the mapping class group.

We have also clarified some important points that were claimed in \cite{nakamura} but were not developed in great depth there. 
We have discussed the relation between the cells of Teichm\"uller space and moduli space, and shown explicitly how the mapping class group acts on cells of Teichm\"uller space.
We have clarified the difference between graphs describing cells in $\cM_{g,n}$ and graphs describing cells in  $\cM_{g,1[n-1]}$ by introducing pole-labelled graphs, and we have related the Euler characteristics of these moduli spaces by counting the graphs of $\cM_{g,n}$ corresponding to each graph of $\cM_{g,n}$.
We have also shown explicitly why the automorphism group of a Nakamura graph is identical to the orbifold group of its cell by explicitly constructing the polytope associated to each graph.

Our description of each cell of Teichm\"uller space consists of two parts: a Nakamura graph $\bar\cG$, and an equivalence class of embeddings $\hat\psi$ of the graph onto a topological surface $\Sigma_{g,n}$. 
While we can construct the boundaries of each cell of $\cM_{g,n}$ by considering the permutation tuple and the defining system of linear equations, we currently do not have an explicit procedure for relating the boundaries of coincident cells in Teichm\"uller space in this description. 
Teichm\"uller space is known to be homeomorphic to an open ball \cite{mcg}. 
One possible aim for such an extension to the Nakamura cell decomposition of Teichm\"uller space would be to show this fact explicitly from the graphs picture.

There exist other descriptions of cell decompositions of moduli space in terms of ribbon graphs.
Strebel quadratic differentials on Riemann surfaces generate embedded ribbon graphs with edge lengths, which also provide a cell decomposition of the moduli space of Riemann surfaces \cite{strebel,Harer,hz,penner,kontsevich}. 
This cell decomposition was claimed in \cite{nakamura} to be less efficient than the Nakamura cell decomposition, as it requires more cells. 
The square of a Giddings-Wolpert differential is a quadratic Strebel differential, but not every Strebel differential has a square root which is a Giddings-Wolpert differential. A future direction of research would be to explore more deeply how these different cell decompositions
are related. Cell decompositions of the {\it compactified} moduli spaces have been considered  from the Strebel point of view \cite{Zuniga}. 
The analogous development in the Nakamura graph picture would be interesting, since it may provide new tools to  
study the  infra-red behaviour of string theory amplitudes which have been of interest recently \cite{Witten,Sam}. 

A concrete problem is to find an explicit  algorithm for homology bases of the Riemann surface corresponding to a given 
Nakamura graph in terms of its combinatoric description. This will be 
useful for using  the Nakamura cell decomposition in connection with 
light-cone string theory scattering amplitude formulae \cite{mand} and for better understanding the action of 
the mapping class group in connection with the metastring \cite{Laurent}.

\vskip2cm 

\begin{centerline}
{ \bf Acknowledgements} 
\end{centerline} 

\vskip1cm 

We thank  Amihay Hanany,  Yang-Hui He and Rodolfo Russo for discussions,  and in particular Laurent Freidel
for collaboration on our closely related earlier paper \cite{fgr}. 
 SR is supported by STFC consolidated grant ST/L000415/1
"String Theory, Gauge Theory \& Duality."

\end{document}